\documentclass[
 article,
 twocolumn,
 groupedaddress,
 showpacs,
 preprintnumbers,
 longbibliography,
 amsmath,
 amsthm,
 amssymb,
 aps,
 prx,
 floatfix,
]{revtex4-1}

\usepackage{graphicx}					
\usepackage{mathdots}
\usepackage{verbatim}					
\usepackage{dcolumn}                    
\usepackage[many]{tcolorbox}
\usepackage[ruled,vlined]{algorithm2e}
\usepackage{color}
\usepackage{fontawesome}
\usepackage{adjustbox}

\usepackage{orcidlink}                     

\newtheorem{prop}{Proposition}
\newtheorem{theorem}{Theorem}

\newcommand{\ds}{\displaystyle}             

\newcommand{\dd}{\mathrm{d}}                
\newcommand{\pd}{\partial}                  

\newcommand{\T}{\mathrm{T}}                 
\newcommand{\R}{\mathrm{R}}                 
\newcommand{\Rf}{\mathrm{Ref}}              
\newcommand{\Rt}{\mathrm{Rot}}              
\newcommand{\Rs}{\mathcal{R}}                 

\DeclareMathOperator{\Tr}{tr}               

\newcommand{\const}{\mathrm{const}}
\newcommand{\ob}{\mathcal{O}}                

\newcommand{\K}{\mathcal{K}}                
\newcommand{\cs}{\mathrm{C.S.}}             

\newcommand{\z}{\zeta}                      

\begin{document}

\title{
Geometry of Almost-Conserved Quantities in Symplectic Maps
\\ Part I. Perturbation Theory
}
\author{T.~Zolkin\,\orcidlink{0000-0002-2274-396X}}
\email{iguanodyn@gmail.com}
\affiliation{Independent Researcher, Chicago, IL}
\author{S.~Nagaitsev\,\orcidlink{0000-0001-6088-4854}}
\affiliation{Brookhaven National Laboratory, Upton, NY 11973}
\affiliation{Old Dominion University, Norfolk, VA 23529}
\author{I.~Morozov\,\orcidlink{0000-0002-1821-7051}}
\affiliation{Synchrotron Radiation Facility ``SKIF'', Koltsovo 630559, Russia}
\affiliation{Novosibirsk State Technical University, Novosibirsk 630073, Russia}
\author{S.~Kladov\,\orcidlink{0000-0002-8005-9373}}
\affiliation{University of Chicago, Chicago, IL 60637}

\date{\today}

\begin{abstract}
Noether's theorem, which connects continuous symmetries to exact
conservation laws, remains one of the most fundamental principles
in physics and dynamical systems.
In this work, we draw a conceptual parallel between two paradigms:
the emergence of exact invariants from continuous symmetries, and
the appearance of approximate invariants from discrete symmetries
associated with reversibility in symplectic maps.
We demonstrate that by constructing approximating functions that
preserve these discrete symmetries order by order, one can systematically uncover hidden structures, closely echoing Noether's
framework.
The resulting functions serve not only as diagnostic tools but also
as compact representations of near-integrable behavior.

The first article establishes the formal foundations of the method.
Using the symmetric form of the map as a flexible test case, we
benchmark the perturbative construction against established
techniques, including the Lie algebra method for twist coefficients.
To resolve the inherent ambiguity in the perturbation series, we
introduce an averaging procedure that naturally leads to a resonant
theory --- capable of treating rational rotation numbers and
small-denominator divergences.
This enables an accurate and structured description of low-order
resonances, including singular and non-singular features in the
quadratic and cubic H\'enon maps.
The approach is systematic, requiring only linear algebra and
integrals of elementary functions, yet it yields results in
striking agreement with both theory and numerical experiment.
We conclude by outlining extensions to more general maps and
discussing implications for stability estimates in practical
systems such as particle accelerators.
\end{abstract}

\maketitle

\section{Introduction}

Among the most profound results in dynamical systems theory, few
have influenced our understanding of long-term behavior as deeply
as the KAM theory
\cite{kolmogorov1954conservation,moser1962invariant,arnol1963small},
Nekhoroshev stability estimates~\cite{nekhoroshev}, and Aubry-Mather
theory~\cite{Siburg2004}.
These cornerstone developments reveal how order and structure can
persist within nonlinear Hamiltonian systems, even under
perturbation --- bridging the divide between integrability and chaos.
Together, they establish a robust framework for understanding
stability, quasi-periodicity, and the delicate onset of transport
and diffusion in complex systems~\cite{Arnold2009_diffusion}.

Equally foundational is Noether's theorem~\cite{Noether1918},
which forged the deep and elegant connection between symmetry
and conservation.
In essence, it states that every continuous symmetry of the action
yields a conserved quantity, a principle that underlies classical
mechanics, field theory, and quantum physics
alike~\cite{Marinho2007,Peskin1995}.
Time translation implies energy conservation;
spatial translation gives linear momentum; rotational symmetry
yields angular momentum.
But more than just a practical tool, Noether's theorem offers a
structural lens --- one that organizes and simplifies complex
systems through their symmetries.

Despite the formal complexity of these theorems, their
implications resonate even in practical settings.
From planetary stability~\cite{siegelmoser1971,arnold2006,Laskar2023}
to modern accelerators~\cite{Giovannozzi1998,Bazzani2019}, we see
that nonlinear systems often exhibit a surprising robustness:
long-term ordered motion, persistence of invariant structures, and
even the survival of near-integrals in strongly perturbed regimes.

This report turns to a related but distinct setting: discrete-time,
symplectic, and reversible dynamical systems/mappings --- a class
dating back to G.D.~Birkhoff~\cite{Birkhoff1917} and explored in
depth by
R.J.~De Vogelaere~\cite{devogelaere1950},
D.C.~Lewis \cite{lewis1961reversible}
and rigorously outlined in a report by
J.A.G. Roberts and G.R.W. Quispel~\cite{roberts1992revers}.
A system is {\it reversible} if there exists an involution in phase
space
--- an operation that undoes itself --- that reverses the direction
of time.
For maps, this means they can be decomposed into the product of two
such involutions.
This structure implies a rich algebra of symmetries: even chaotic
reversible maps possess an infinite discrete symmetry group generated
by compositions of involutions and map iterates.

Here, we ask: can these symmetries help uncover approximate
invariants in chaotic or near-integrable systems --- especially
in settings where traditional integrals are no longer exact? 
For an integrable map of the plane $\mathrm{M}$, the invariant
function satisfies
\[
    \K[\mathrm{M}(p,q)]-\K[p,q] = 0
\]
for any values of $(p,q)\in\mathbb{R}^2$, and is itself invariant
not only under the map, but both involutions comprising it.
This property, famously utilized by McMillan in constructing an
integrable map, extends beyond integrable cases: even in chaotic
systems, invariant structures tend to reflect the underlying
symmetries of the map.
These symmetries aid in identifying symmetric fixed points,
$n$-cycles, and organizing bifurcations, see, for example,
\cite{roberts1992revers, zolkinHenonSet}.

Building on this insight, we propose a perturbative framework that
generalizes this idea.
Rather than demanding exact invariance, we construct an approximate
invariant
\[
    \K^{(n)}[\mathrm{T}(p,q)]-\K^{(n)}[p,q] = \ob(\epsilon^{n+1})
\]
where $\mathrm{T}$ is a general symplectic, possibly chaotic map.
This deceptively simple condition --- requiring that the function
is preserved up to a given order --- proves to be a powerful
principle.
It allows us to:
(i) match twist coefficients (expansion of rotation number on action
variable), consistent with Lie algebra techniques;
(ii) recover normal forms for low-order nonlinear resonances, that
perfectly align with numerical experiments;
(iii) reconstruct global invariant curves for exactly integrable
systems; and
(iv) approximate the geometry of island chains and low-order KAM
structures.
The method itself is systematic and relies only on solving linear
systems and, when necessary, integrating powers of trigonometric
functions to handle small denominator problems.
This makes it not only analytically tractable, but also
geometrically illuminating.

A particularly striking outcome is that approximate invariants
derived in this way naturally inherit the discrete symmetries of
the system --- mirroring the role continuous symmetries play in
Noether's theorem.
Symmetry, once again, becomes the organizing principle behind
conservation --- even if only approximately.

As a benchmark, we provide a next-order extension of the
Courant-Snyder~\cite{courant1958theory} invariant used in
accelerator physics, incorporating leading nonlinear terms.
Numerous examples are explored, including real-world applications.

We believe that this approach --- simple in its construction yet
rich in implications --- offers a promising tool for probing
structures in nonlinear maps and opens the door to deeper insights
into the resilience of order in chaotic systems.

\subsection{Motivation and preliminary remarks}

In the study of dynamical systems --- particularly within
symplectic and Hamiltonian mechanics --- Lie algebras and Lie
transformations serve as powerful tools for computing {\it
action-angle variables} and {\it twist coefficients} $\tau_i$,
which characterize how the angle variable $\theta$ evolves with
the action $J$, i.e., how the nonlinear {\it frequency} $\nu$
shifts with amplitude.
The central idea is to transform the Hamiltonian (or map) into
a canonical form
\cite{MichelottiLeo1995,Berz1999,Forest1998,morozov2017dynamical}:
\[
\mathcal{H}[J,\theta;t] =   \nu_0\,J + \frac{\tau_0}{2!}\,J^2 +
                \frac{\tau_1}{3!}\,J^3 + \ldots,
\]
in which the Hamiltonian depends solely on the action $J$,
achieved through a sequence of Lie transformations.
This formalism systematically eliminates angle dependence order
by order.

In accelerator physics, Lie algebra methods have been successfully
employed to optimize lattice designs by removing {\it singular}
resonances~\cite{bengtsson1997,Bingfeng2023,Bingfeng2024} --- those
for which the detuning $\dd\nu/\dd J$ diverges at $J=0$.
Additionally, determining the shift in frequency (or
{\it rotation number}/{\it betatron tune} in the discrete-time
case) as a function of amplitude is crucial for estimating
thresholds of collective instabilities and Landau damping
\cite{Landau1965,Malmberg1964} required for their mitigation
\cite{Gareyte1997,Shiltsev2017}.

These transformations effectively straighten the motion: in the
transformed coordinates, trajectories trace circles, the action
remains invariant, and the angle increases linearly in time,
governed by an amplitude-dependent frequency.
However, this elegant behavior holds only in regions of regular,
bounded motion --- and this is where the complications begin.

Typically, integrable systems possess {\it separatrices} ---
boundaries dividing regions of qualitatively different motion.
These are formed by stable and unstable manifolds of hyperbolic
(unstable) fixed points and/or periodic orbits ($n$-cycles).
As a result, any action-angle formulation is inherently local,
valid only within regions enclosed by separatrices, and fails to
describe the global phase space structure.
The issue is even more severe in chaotic systems, where the very
concept of a simply connected domain of stability may break down.
In such systems, even an infinitesimally small amplitude motion may
be interrupted by chains of islands, arising from the breakdown
of invariant tori~\cite{lichtenberg2013regular}.

This motivates the development of perturbative methods that can
complement the Lie algebra approach.
Unlike Lie techniques, which focus on infinitesimal neighborhoods
around stable fixed points, perturbative methods aim to provide
meaningful estimates beyond the separatrix, potentially navigating
through island chains and capturing aspects of the broader dynamics.

To illustrate these ideas and set the stage for further discussion,
we begin with the case of continuous-time dynamics.
Although our eventual perturbative construction will differ in
specifics, this preliminary discussion highlights the challenges
and outlines alternative strategies.
As examples, consider the three following classical Hamiltonians:
\[
\begin{array}{l}
\ds\mathcal{H}_1[p,q;t] = \frac{p^2}{2} + \frac{q^2}{2} - \frac{q^4}{4},   \\[0.3cm]
\ds\mathcal{H}_2[p,q;t] = \frac{p^2}{2} + \frac{q^2}{2} -
    \frac{3}{2\,\sqrt{2}}\,\frac{q^3}{3} +
    \frac{1}{4}\,\frac{q^4}{4},                                     \\[0.3cm]
\ds\mathcal{H}_3[p,q;t] = \frac{p^2}{2} + \frac{q^2}{2}
    + \sum_{i=0}^{\infty}
    \frac{\tau_i}{(i+2)!}
    \left(\frac{p^2+q^2}{2}\right)^{i+2},
\end{array}
\]
each representing different nonlinear oscillators.

The first system is a classical harmonic oscillator with a cubic
nonlinearity, i.e., it includes a restoring force of the form
$F(q) = -q + q^3$.
The second system, though perhaps less familiar in this form, can
be derived from the Duffing oscillator~\cite{duffing1918erzwungene}:
\[
H_\mathrm{Duff.}[p,Q;\mathrm{t}] =
    \frac{p^2}{2} - \frac{Q^2}{2} + \frac{Q^4}{4},
\]
by rescaling time via $\mathrm{t} = t/\sqrt{2}$, with the corresponding
shift of the origin according to $Q=(q-1)/\sqrt{2}$ and
$H_\mathrm{Duff.}= \mathcal{H}_2 - \frac{1}{4}$.

The third system is defined for
\[
    J \leq J_\mathrm{max} = \frac{2^{3/2}}{3\,\pi}
\]
where $J$ is the canonical action variable, proportional to the
area enclosed by an invariant torus in phase space
(i.e., a level set of constant $\mathcal{H}_3[p,q]=\const$):
\[
    J = \frac{1}{2\,\pi}\oint p\,\dd q = \frac{p^2+q^2}{2}.
\]
Accordingly, the Hamiltonian is naturally expressed in terms of
action-angle variables and takes the form
\[
    \mathcal{H}_3[J,\theta;t] = \nu_0J +
    \sum_{i=0}^{\infty}
    \frac{\tau_i}{(i+2)!}J^{i+2},
    \qquad\qquad
    \nu_0 = 1.
\]
It is independent of the angle variable $\theta$, reflecting an
exact continuous symmetry and, by Noether's Theorem, guaranteeing
conservation of the action  $J$.

The twist coefficients $\tau_i$ describe the frequency's
dependence on the action via a power series:
\[
    \nu = \frac{\dd\theta}{\dd t} = \frac{\pd\mathcal{H}}{\pd J} =
    \nu_0 + \sum_{i=0}^{\infty}\frac{\tau_i}{(i+1)!}J^{i+1}
\]
and for this example we choose them to match the series
\[
\nu(J) = 1 -   \frac{3}{4}\,J - \frac{51}{64}\,J^2 -
            \frac{375}{256}\,J^3 - \frac{53445}{16384}\,J^4 +
            \ob(J^5)
\]
which arises from an implicit dependence given by:
\begin{equation}
\label{math:NuJ123}    
\begin{array}{l}
\ds\nu(\kappa) =\,\,\,
\frac{\pi}{2}\,\frac{\mathrm{K}^{-1}[\kappa]}
                    {\sqrt{1+\kappa^2}},                    \\[0.35cm]
\ds  J(\kappa) = 
\frac{4}{3\,\pi}\frac{1}{\sqrt{1+\kappa^2}}
\left(
    \mathrm{E}[\kappa] - 
    \frac{1-\kappa^2}{1+\kappa^2}\mathrm{K}[\kappa]
\right),
\end{array}
\end{equation}
where $\mathrm{K}[\kappa]$ and $\mathrm{E}[\kappa]$ are the
{\it  complete elliptic integrals of the first and second kind},
respectively, and $\kappa\in[0,1]$ is the {\it elliptic modulus}.

\newpage
\begin{figure}[t!]
    \includegraphics[width=\linewidth]{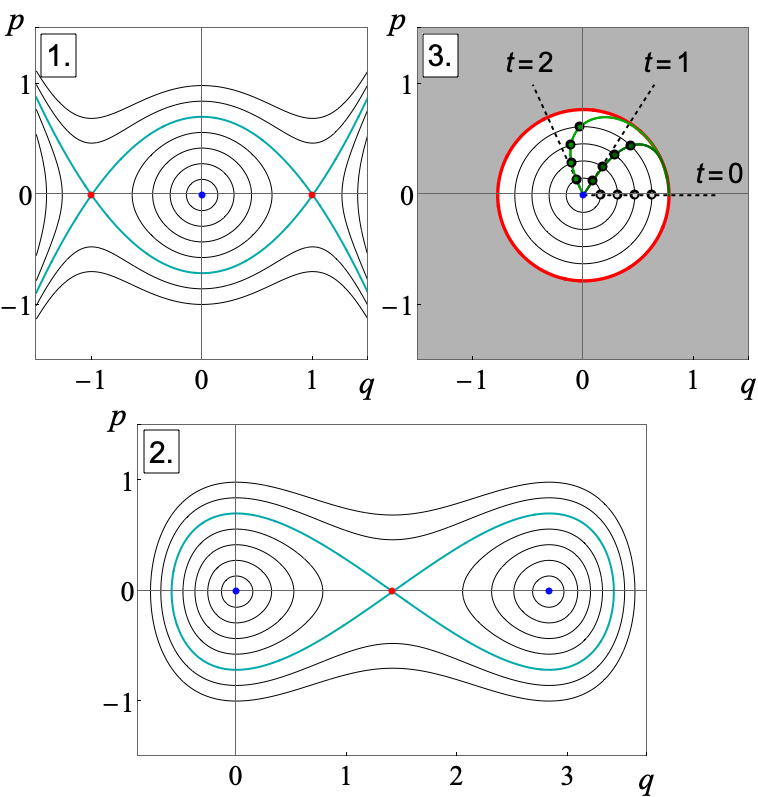}
    \caption{\label{fig:IntroPS123}
    Constant level sets of Hamiltonians $\mathcal{H}_{1,2,3}[p,q]$.
    Blue and red points mark stable ({\it elliptic centers}) and
    unstable ({\it hyperbolic saddles}) stationary solutions,
    respectively.
    Cyan curves indicate (homo-)heteroclinic connections forming
    separatrices between different types of motion.
    In the third plot, the gray region denotes the domain where
    the Hamiltonian is not defined.
    Additionally, the last plot illustrates a sequence of points
    selected on invariant curves and along the line $q=0$ at $t=0$,
    along with their images under the dynamics at $t=1$ and $t=2$.
    Green lines show the forward iterates of an entire segment of
    the $p>0$ half-line.
    }
\end{figure}
\begin{figure}[t!]
    \includegraphics[width=\linewidth]{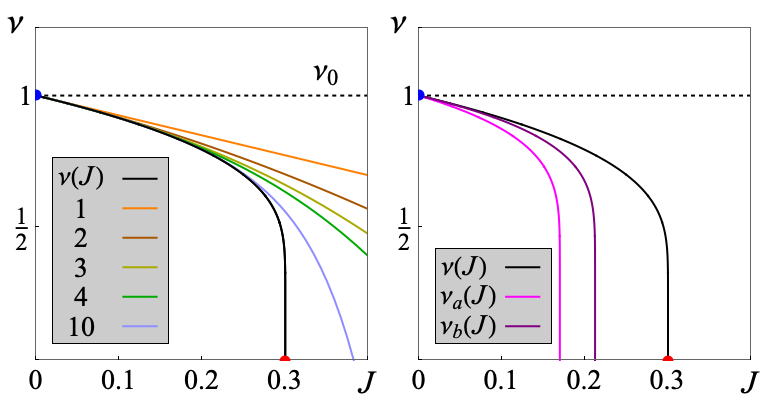}
    \caption{\label{fig:IntroNuJV}
    Frequency of oscillations $\nu(J)$ as a function of the action
    variable for the model Hamiltonians $\mathcal{H}_{1,2,3}$
    (black curve).
    The left plot shows power series approximations of the form
    $\nu_n(J) \approx \nu_0 + \tau_0\,J + \ldots + \tau_{n-1}\,J^n$,
    while the right plot compares these with approximations obtained
    from Hamiltonians featuring cubic potentials.
    }\vspace{-0.5cm}
\end{figure}

Interestingly enough, while we did some ``magic'' by purposefully
matching action angle variables of Hamiltonians $\mathcal{H}_1$ to
$\mathcal{H}_3$ by construction, the Hamiltonian $\mathcal{H}_2$
also has exactly the same $\nu(J)$ dependence around the stationary
point at the origin, given by Eqs.~(\ref{math:NuJ123}).
Plots 1 through 3 in Fig.~\ref{fig:IntroPS123}
provide phase space diagrams with level sets of corresponding
Hamiltonians, and we see that despite exactly the same description of
dynamics within the simply connected region, the complete picture
differs a lot.
At large amplitudes, in case (1.) we have unstable motion, in case
(2.) stable orbits rounding the origin, and in case (3.) the system
is not defined.
Moreover, even at the boundary of motions, in the first case we have
a separatrix made of a {\it heteroclinc connection} between two
different saddles, two {\it homoclinic orbits} attached to a single
unstable solution in case (2.), and a boundary completely made
of fixed points in the last case.
This example illustrates how information is inevitably lost beyond
$J\geq J_\mathrm{max}$, even if all twist coefficients $\tau_i$ are
known and/or a closed-form expression for $\nu(J)$ is available.
The left plot in Fig.~\ref{fig:IntroNuJV} shows the dependence of
$\nu(J)$, compared to its truncated power series approximations.

\newpage
With that in mind, while the transformation to action-angle
variables effectively ``blinds'' us beyond the separatrix, the
original Hamiltonian equations of motion,
\[
\frac{\dd q}{\dd t} = \frac{\pd \mathcal{H}}{\pd p}
\quad\text{and}\qquad
\frac{\dd p}{\dd t} =-\frac{\pd \mathcal{H}}{\pd q},
\]
still contain full information about the system, including the
globally conserved values of $\mathcal{H}$.

Now, focusing on case 2, suppose we do not know the exact
invariant $\mathcal{H}_2[p,q]$, but we do have access to the
equations of motion.
Further, imagine we've developed a perturbative approach capable
of reconstructing the invariant from the equations of motion in
the form of a polynomial.
Then, in zeroth order (often referred to as a {\it linearization}),
we would recover the harmonic oscillator:
\[
\mathcal{H}_0[p,q;t] = \frac{p^2}{2} + \frac{q^2}{2}.
\]
While this oversimplified approximation cannot predict any
nonlinear phenomena --- such as detuning or even the presence
(let alone the location) of a separatrix --- it still provides
the correct {\it bare} frequency $\nu_0 = \nu(0)$.

In the next order, we can imagine that our hypothetical
perturbation theory successfully recovers the cubic term
in the potential energy
(corresponding to the quadratic term in the restoring force):
\[
\mathcal{H}_{2.a}[p,q;t] = \frac{p^2}{2} + \frac{q^2}{2} -
    \frac{3}{2\,\sqrt{2}}\,\frac{q^3}{3}.
\]
The right plot in Fig.~\ref{fig:IntroNuJV} illustrates the
corresponding frequency-action dependence $\nu_a(J)$.
What is immediately apparent is that although the prediction
for the location of the separatrix is underestimated by
almost a factor of two, the behavior of $\nu_a(J)$ captures
the qualitative aspect far better than the traditional power
series shown in the left plot.
However, one key issue remains: the approximate Hamiltonian
$\mathcal{H}_{2.a}[p,q;t]$ fails to reproduce the correct
twist coefficient $\tau_0=-3/4$, yielding instead an incorrect
value of $-15/16$.

\begin{figure}[t!]
    \includegraphics[width=\linewidth]{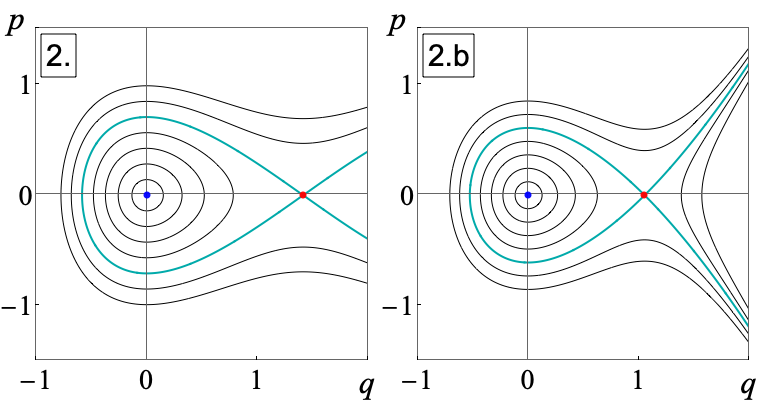}
    \caption{\label{fig:IntroPS22b}
    Constant level sets of Hamiltonians $\mathcal{H}_2$ and
    $\mathcal{H}_{2.b}$.
    }
\end{figure}

Thus, before proceeding to recover the next order term, we can attempt
to refine our approximation within the same order by modifying it:
\[
\mathcal{H}_{2.b}[p,q;t] = \frac{p^2}{2} + \frac{q^2}{2} -
    \frac{3}{\sqrt{10}}\,\frac{q^3}{3}
\]
so that it restores the correct value of $\tau_0=-3/4$.
Looking at the behavior of the refined frequency $\nu_{2.b}(J)$
shown in Fig.~\ref{fig:IntroNuJV}, we see that although it still
deviates, the estimated location of the separatrix now achieves
accuracy comparable in absolute value with the tenth order of the
traditional power series --- and once again provides much better
qualitative agreement.
Additionally, Fig.~\ref{fig:IntroPS22b} presents a direct
comparison of the level sets defined by the exact Hamiltonian
$\mathcal{H}_2$ and its approximation $\mathcal{H}_{2.b}$.
While the approximation still cannot describe the dynamics beyond
the separatrix, it captures its presence and essential qualitative
features of the motion within its vicinity remarkably well.

We now shift focus to the central subject of this article:
{\it symplectic mappings} of the plane~\cite{arnold1989ergodic},
defined by a pair of coupled equations
\[
\begin{array}{l}
q' = f_1(q,p),  \\[0.25cm]
p' = f_2(q,p),
\end{array}
\]
where we will use the prime symbol ($'$) throughout to denote the
image under one application of the map.
Symplectic maps preserve phase space volume and the underlying
geometric structure, mapping infinitesimal elements to others of
equal measure.
For maps of the plane, this symplectic condition is equivalent to
the conservation of phase space area.

One way such maps arise is through discretizing the continuous-time
flow of a Hamiltonian system with one degree of freedom at
equidistant time intervals.
These mappings are necessarily integrable, possessing an exact
integral of motion (given by any function of the generating
Hamiltonian itself).
An example of such a discretization, applied to the Hamiltonian
$\mathcal{H}_3$
is shown in plot (3.) of Fig.~\ref{fig:IntroPS123}.
The resulting map serves as an example of a nonlinear, integrable
{\it twist map}~\cite{arnold1989ergodic,moser1962invariant}.

\begin{figure}[t!]
    \includegraphics[width=\linewidth]{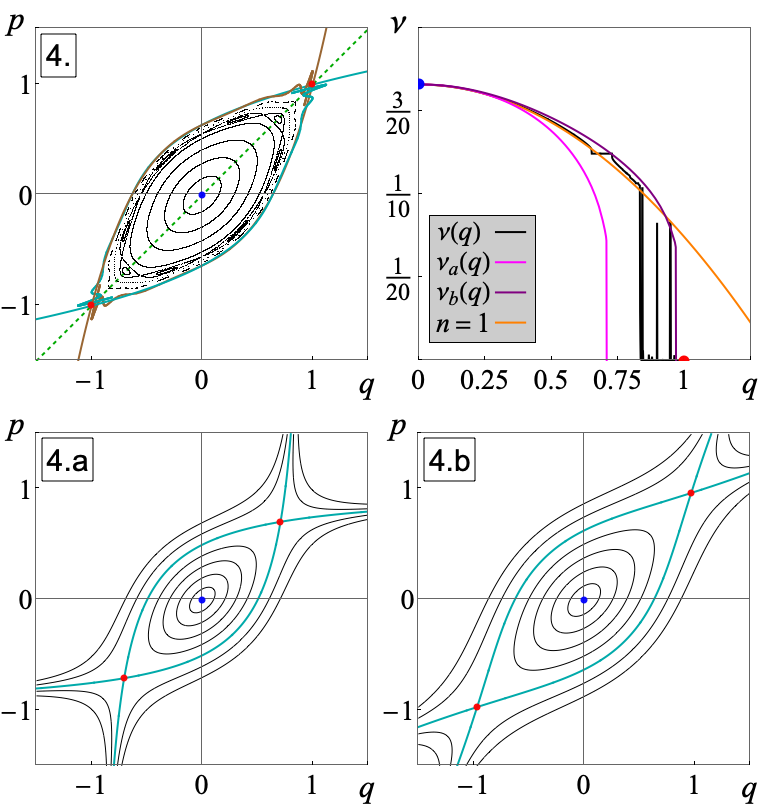}
    \caption{\label{fig:IntroPS4all}
    Top left: Some invariant structures of the cubic H\'enon map
    with $f(p) = p + p^3$, including surviving KAM circles (black),
    a stable fixed point (blue), unstable fixed points (red), and
    their associated stable and unstable manifolds (brown and cyan),
    computed via continuation of linearized eigenvectors.
    The dashed green line indicates the first symmetry, $p=q$.
    Top right: Rotation number $\nu(q)$ along the symmetry line
    (black curve), alongside several approximations:
    the quadratic truncation of the traditional power series
    $\nu(q) = \nu_0 + (2\,\pi\sqrt{3})^{-1}q^2$, the integrable
    McMillan-based approximation $\nu_{4.a}(q)$, and a refined
    version derived from the averaged invariant of the same order
    $\nu_{4.b}(q)$.
    Bottom: Level sets of the approximated invariants
    $\K_{4.a}[p,q]$ and $\K_{4.b}[p,q]$.
    }
\end{figure}

Another route to symplectic maps involves taking a
{\it stroboscopic Poincar\'e section} at times $t_k = k\,T$ of a
non-autonomous Hamiltonian system with one degree of freedom that
is subject to a time-periodic external force (with period $T$).
Alternatively, they can arise from regular Poincar\'e sections of
autonomous Hamiltonians with two degrees of
freedom~\cite{lichtenberg2013regular}.
Famous examples include the chaotic Chirikov standard map
\cite{chirikov1969research,chirikov1979universal} and the
H\'enon–Heiles potential~\cite{henon1964applicability}.
In contrast to the previous (integrable) case, these systems
generally exhibit chaos, with integrability being the exception
rather than the rule~\cite{mcmillan1971problem,suris1989integrable,
brown1993,ZKN2023PolI,ZKN2024PolII}.

Before diving into the construction and analysis of our
perturbative method, let’s preview what's to come through one
more illustrative example, the chaotic H\'enon map
\cite{henon1969numerical,sterling1999homoclinic,dullin2000twistless,
dulin2000henon,zolkinHenonSet} with a cubic force $f(p) = p + p^3$:
\[
\begin{array}{l}
q' = p,  \\[0.25cm]
p' =-q + f(p).
\end{array}
\]
Much like the previously discussed Hamiltonian
$\mathcal{H}_1[p,q;t]$, this map possesses a stable fixed point
at the origin, as well as two unstable fixed points --- this
time located along the first symmetry line $p=q$.
However, in this case, the stable and unstable manifolds no longer
form a heteroclinic connection.
Instead, they form a {\it heteroclinic intersection} or {\it tangle}
(see the cyan and brown curves in the top-left plot of
Fig.~\ref{fig:IntroPS4all}).
As a result, the map is not integrable, and an {\it exact
invariant} $\mathcal{K}_4[p,q]$ satisfying
\[
    \mathcal{K}_4[p',q'] - \mathcal{K}_4[p,q] = 0,
\]
for arbitrary $q$ and $p$ simply does not exist.
The action variable can no longer be meaningfully defined ---
especially for large-amplitude dynamics.
The rotation number as a function of position becomes a non-smooth
function, exhibiting flat mode-locked regions (akin to the
well-known Arnold tongues in the standard circle map
\cite{Arnold2009,BoylandCircle,KuznetsovCircle} during parametric
resonance).
For even larger amplitudes, not only smoothness but also continuity
of $\nu(J)$ breaks down, as the system transitions to fully chaotic
motion with mixing~\cite{sterling1999homoclinic}.
Nevertheless, alongside fixed points, $n$-cycles, and their
attached manifolds, the system still retains invariant structures,
notably KAM tori (shown in black), which represent surviving regular
orbits with irrational values of $\nu$.
So our goal is to construct, using only the equations of motion,
an approximate function satisfying:
\[
\mathcal{K}_4^{(n)}[p',q'] - \mathcal{K}_4^{(n)}[p,q] =
    \ob(\epsilon^{n+1}).
\]

In a preceding series of articles
\cite{zolkin2024MCdynamics,zolkin2024MCdynamicsIII},
we explored connections between {\it integrable McMillan maps}
\cite{mcmillan1971problem,IR2001I,IR2002II,IR2003III}
and the chaotic H\'enon map (and
more broadly, {\it typical} transformations in the {\it McMillan
form}).
These approximations reproduce the dynamics up to the first order in
$J$, and admit an approximate invariant:
\[
\mathcal{K}_{4.a}^{(2)}[p,q] = p^2 - p\,q + q^2 - p^2\,q^2
\]
corresponding to a map whose equations of motion agree with the
H\'enon system up to $\ob(p^5)$
\[
\begin{array}{l}
q' = p,  \\[0.2cm]
\ds p' =-q + \frac{q}{1-q^2} = -q + p + p^3 + \ldots.
\end{array}
\]
Notably, with the same nonlinear tune shift $\tau_0$ at the origin,
see the magenta curve labeled $\nu_{a}$ in the top-right plot of
Fig.~\ref{fig:IntroPS4all}.

In the upcoming discussion, we address two important questions.
(i) The first concerns how to systematically proceed to higher
orders in such a way that, at a minimum, the twist coefficients
$\tau_i$ are matched precisely.
As we will discover, this question admits infinitely many solutions.
The source of this indeterminacy lies in the fact that the invariant
is only known up to an arbitrary function of itself, and, is
directly related to the idea that the local dynamics within a simply
connected region (or within an infinitesimally small neighborhood of
a fixed point, in the presence of chaos) can correspond to
drastically different global systems.
(ii) Although in our previous Hamiltonian examples a refinement
was required to match the twist coefficient, in the current case
--- where $\tau_0$ is already correctly reproduced --- some
further adjustment can still be made.
Such refinement allows us to select among the infinitely many
undetermined higher-order terms.
For our case of the cubic H\'enon map, an improved
second-order approximate invariant takes the form:
\[
\mathcal{K}_{4.b}^{(2)}[p,q] = p^2 - p\,q + q^2 - p^2\,q^2
+ \frac{7}{5}(p^2 - p\,q + q^2)^2.
\]
The bottom row in Fig.~\ref{fig:IntroPS4all} shows the level sets
of both $\mathcal{K}_{4.a}^{(2)}[p,q]$ and
$\mathcal{K}_{4.b}^{(2)}[p,q]$, while the top-right plot compares
their respective $\nu(q)$ dependencies.
For reference, it also includes the truncated power series
expansion $\nu(q) = \nu_0 + (2\,\pi\sqrt{3})^{-1}\,q^2$ (orange),
that can be derived via Lie transformations or expansions
provided in~\cite{zolkin2024MCdynamics}.
Among the three approximations, the refined invariant
$\mathcal{K}_{4.b}^{(2)}[p,q]$ gives the best estimate for the
location of the unstable fixed point (marked with a red dot) and
provides the best overall qualitative agreement.
Before moving forward, it is worth noting that in systems exhibiting
chaos, the size of the simply connected region obtained via
approximations shrinks rapidly as one moves to higher orders.
Consequently, low-order approximations often carry the most
meaningful insights, capturing essential qualitative features
of the system's behavior --- features that tend to persist across
a wide range of parameters and remain robust even when higher-order
corrections become unreliable or analytically intractable.

\vspace{-0.5cm}
\subsection{Article Structure}

\vspace{-0.4cm}
Taking these considerations into account, we present our findings
across two complementary articles.
In the first article we focus on establishing the perturbative
method and its technical foundations (Section~\ref{sec:PerT}).
As an illustrative example, we use a special {\it McMillan form}
of the map, which offers great flexibility across different
nonlinear mappings, while remaining symplectic and reversible.
Our primary goal here is to verify convergence in an
infinitesimally small neighborhood of the fixed point.
This is done by comparing twist coefficients obtained via our
method with those derived from the well-established Lie algebra
technique (Subsection~\ref{sec:Action}).
To overcome the earlier-mentioned indeterminacy, we introduce
the {\it averaging procedure} (Subsection~\ref{sec:Averaging}).
Interestingly, this transforms our perturbative approach into a
{\it resonant theory}, i.e., one capable of handling situations
where the rotation number is rational (commensurate with unity).
This advancement allows us to handle diverging denominators
(Subsection~\ref{sec:Removal}) and explains the rich structure
of small-amplitude oscillations.
This resonant capability is essential: in systems exhibiting
nonlinear resonance, even infinitesimal amplitudes can produce
behavior that contradicts traditional linearization techniques.
We illustrate this by exploring both singular and nonsingular
low-order resonances in the quadratic and cubic H\'enon maps.

Finally, to complete the technical description, Section
\ref{sec:Generalize} demonstrates how the method can be extended
to general map forms and outlines a path toward applications in
higher-dimensional systems.
In the closing discussion, Section~\ref{sec:Conclusion}, we
explore how our findings may offer new insights into the
longstanding question of the size of stable KAM regions,
particularly illuminating the remarkable stability observed
in particle accelerators despite dramatic variations in their
nonlinear optics designs.

In the second article, we turn to potential applications of the
perturbative method by exploring large-amplitude dynamics.
Here, we put our approach to a more demanding test: attempting to
recover the global invariant in integrable mappings, where such
invariants are known to exist.
We demonstrate that, for McMillan integrable mappings with finite
polynomial invariants, our perturbation theory converges exactly
at the expected order.
For mappings with infinite series invariants (such as the Suris trigonometric and exponential maps), our method appears to yield
almost everywhere convergence, allowing the invariant to be
recursively constructed.
Having passed all critical benchmarks --- accurate recovery of
twist coefficients, correct treatment of rational rotation numbers,
and convergence to known invariants --- we move to potential
applications.
In particular, we attempt to reconstruct the behavior of the
rotation number and a global invariant that captures dominant
island structures.
This analysis is carried out in an extended space of dynamical
variables and mapping parameters, allowing for much broader
coverage of possible regimes and avoiding the pitfall of being
confined to specific phase space portraits.
We conclude by comparing our method to the recently introduced
Square Matrix (SM) technique~\cite{Hua17, Hua2023, square}.
As an example of a real-life application, we demonstrate how a
single, moderately complex function (representing a low-order
approximate invariant) can be used to reconstruct single particle
dynamics across a range of seemingly unrelated accelerator
facilities.
Remarkably, this single expression captures essential nonlinear
behavior in systems as different as the resonant extraction for
Mu2e at the g-2 experiment, the Main Injector ring and the Integrable
Optics Test Accelerator (IOTA) at FermiLab.
This unexpected unification underscores the power of low-order
approximations in revealing shared structural dynamics across
diverse machines.

Because this article benchmarks a broader body of prior research,
it is important to mention several relevant auxiliary results.
Articles~\cite{zolkin2017rotation, nagaitsev2020betatron,
mitchell2021extracting} contain supporting material on the Danilov
theorem, which helps bridge our perturbation theory with standard
action-angle variables.
The series~\cite{zolkin2024MCdynamics, zolkin2024MCdynamicsII,
zolkin2024MCdynamicsIII} provides a detailed treatment of low-order
integrable approximations.
Article~\cite{zolkinHenonSet} offers a comprehensive explanation
of the isochronous diagrams used in the second part of this study.
Finally, we acknowledge the foundational review of reversibility
in dynamical systems by Roberts and Quispel
\cite{roberts1992revers}.

\newpage

\newpage
\section{\label{sec:PerT}Perturbation theory}

In this section, we develop a method to construct an approximate
invariant of motion for (quasi-) nearly integrable orbits
governed by a symplectic map, $\T:\z\mapsto\z'$, defined on the
plane $\z=(q,p)\in\mathbb{R}^2$, and which can be expressed in
the so called {\it McMillan form}~\cite{
mcmillan1971problem,
suris1989integrable,
roberts1992revers,
IR2001I,IR2002II,IR2003III,
ZKN2023PolI,ZKN2024PolII,
zolkin2024MCdynamics,zolkin2024MCdynamicsIII}:
\begin{equation}
\label{math:Ms}
\begin{array}{ll}
\T: \qquad  & q' = p,           \\[0.25cm]
		& p' =-q + f(p),
\end{array}
\end{equation}
where $f(p)$ is a $C^\infty$ function referred to as the
{\it force function}, or simply {\it force}.
While $\T$ may not exhibit global quasi-integrability, we assume
the existence of invariant KAM tori in the neighborhood of a fixed
point~\cite{
kolmogorov1954conservation,moser1962invariant,arnol1963small}.

The map is reversible~\cite{
devogelaere1950,
lewis1961reversible,
roberts1992revers,
zolkinHenonSet},
and therefore, along with its inverse, it can be expressed as a
composition of two involutory, anti-area-preserving transformations,
called {\it reversors}:
\[
    \T = \R_2 \circ \R_1,
    \qquad
    \T^{-1} = \R_1 \circ \R_2,
    \qquad
    \R_{1,2}^2 = \mathrm{Id},
    \]
where
\[
\begin{array}{ll}
\R_1:   & q' = p,                   \\[0.25cm]
        & p' = q,
\end{array}
\qquad\qquad
\begin{array}{ll}
\R_2:   & q' = q,                   \\[0.25cm]
	& p' =-p + f(q).
\end{array}
\]
The first reversor, $\R_1 = \Rf(\pi/4)$, corresponds to a linear
reflection with respect to the main diagonal, where
\[
\Rf(\theta) = 
\begin{bmatrix}
    \cos(2\theta) & \sin(2\theta)   \\[0.25cm]
    \sin(2\theta) &-\cos(2\theta)
\end{bmatrix}.
\]
The second reversor performs a vertical reflection about the
line $p = f(q)/2$.
These lines, known as {\it symmetry lines} or simply
{\it symmetries} of $\T$, correspond to the fixed-point sets of
the reversors:
\[
l_1 = \mathrm{Fix}\,(\R_1):\,\,
    p = q,                  \qquad
l_2 = \mathrm{Fix}\,(\R_2):\,\,
    p = f(q)/2.
\]

If any point of an orbit belongs to
$\mathcal{M}_n = \mathrm{Fix}\,(\T^n\R_{1,2})$, the orbit is
termed {\it symmetric}.
Otherwise, it is referred to as {\it asymmetric}.
Points that belong to
$\mathcal{M}_{m,n} = \mathcal{M}_m\cap\mathcal{M}_n$
are called {\it doubly symmetric}.
It follows that~\cite{devogelaere1950,lewis1961reversible}:
(i) every doubly symmetric point is periodic under $\T$, and
(ii) every symmetric periodic point is doubly symmetric.
In particular, all fixed points of the system,
$\z^{(1)} = \T\,  \z^{(1)}$, and 2-cycles,
$\z^{(2)} = \T^2\,\z^{(2)}$, are given as intersections of
$l_1$ and $l_2^{\pm1}$, respectively:
\[
\z^{(1)}:\,\left\{
\begin{array}{l}
	p = q,             \\[0.25cm]
	p = f(q)/2,
\end{array}
\right.
\quad\text{and}\quad
\z^{(2)}:\,\left\{
\begin{array}{l}
	q = f(p)/2,        \\[0.25cm]
	p = f(q)/2.
\end{array}
\right.
\]

In the following derivations, we will assume, without loss of
generality, that at least one fixed point is located at the origin.
This can always be achieved by a constant shift of coordinates
$(q,p)\rightarrow(q,p) + \lambda\,(1,1)$, where $\lambda$ is chosen
such that $f(0)=0$.
The linear stability of this fixed point is defined by the trace
of its Jacobian matrix:
\[
\dd\T(\z) = \begin{bmatrix}
		\frac{\pd q'}{\pd q} & \frac{\pd q'}{\pd p} \\[0.15cm]
		\frac{\pd p'}{\pd q} & \frac{\pd p'}{\pd p}
	\end{bmatrix}
      = \begin{bmatrix}
		0 & 1 \\[0.25cm]
		-1& \pd_p f(p)
	\end{bmatrix},
\]
where the condition $|\Tr\dd\T(\mathbf{0})|<2$, gives the stability
criterion:
\[
    |a|<2, \qquad\qquad a = \pd_p f(0),
\]
with $a$ being related to the {\it bare}/{\it unperturbed rotation
number}:
\[
    \nu_0 = \arccos[a/2]/(2\,\pi).
\]

Next, any set of points $\Gamma$ is called {\it invariant} under
$\T$, if:
\begin{equation}
\label{math:TG}
    \T\,\Gamma = \Gamma.
\end{equation}
If the mapping is integrable, it should possess an {\it exact
invariant of motion} $\K[p,q]$, such that for any point of the
plane $\z\in\mathbb{R}^2$ and its image $\z'$:
\begin{equation}
\label{math:KK'}    
    \K[\z'] - \K[\z] = 0.
\end{equation}
As noted by E. McMillan~\cite{mcmillan1971problem}, the constant
level set of such an invariant is not only invariant under $\T$,
Eq.~(\ref{math:TG}), but must also be invariant under both
reflections:
\begin{equation}
\label{math:RR'} 
    \K[\R_1\z] - \K[\z] = 0,
    \qquad\qquad
    \K[\R_2\z] - \K[\z] = 0.
\end{equation}
Thus, in general, the constant level set of the invariant of a
reversible mapping, $\K[p,q]=\const$, consists of symmetric
invariant subsets and pairs of asymmetric invariant subsets
\cite{roberts1992revers}.
This makes the invariance condition (\ref{math:KK'}) more general
than (\ref{math:TG}), as the entire union $\K[p,q]=\const$ must
necessarily satisfy (\ref{math:RR'}).

To develop a perturbation theory, we introduce a small positive
parameter, $\epsilon$, to characterize the amplitude of
oscillations.
This can be achieved by performing a change of variables,
$(q,p) \rightarrow \epsilon\,(q,p)$.
Under this transformation, the map~(\ref{math:Ms}) retains its
original form, with the force function modified as follows:
\[
f(p) \rightarrow \frac{f(\epsilon\,p)}{\epsilon} =
    a\,p + \epsilon\, b\,p^2
         + \epsilon^2 c\,p^3
         + \epsilon^3 d\,p^4
         + \ldots,
\]
where $f(p)$ is expanded into a power series
\[
	b \equiv \frac{\pd_p^2 f(0)}{2!},\quad
	c \equiv \frac{\pd_p^3 f(0)}{3!},\quad
	d \equiv \frac{\pd_p^4 f(0)}{4!},\quad
	e \equiv \frac{\pd_p^5 f(0)}{5!}.
\]

Finally, as we will observe, enforcing the first symmetry $p = q$
is essential, prompting the introduction of the following notations:
\[
\Sigma = p + q,
\qquad\qquad
\Pi = p\,q,
\]
along with their combination
\[
\qquad\qquad
\cs = \Sigma^2 - (2+a)\,\Pi = p^2 - a\,p\,q + q^2,
\]
corresponding to the zeroth-order (linearized) invariant, commonly
known as the {\it Courant-Snyder invariant}~\cite{courant1958theory},
and playing a significant role in the expansion.

\subsection{\label{sec:Construct}Approximate invariant of motion}

After introducing $\epsilon$, we construct a perturbation theory
in which, at each order $n$, we seek an {\it approximate invariant}, 
assumed to take the form of a power series 
\[
\K^{(n)}[p,q] = \K_0 + \epsilon\,\K_1 + \epsilon^2\,\K_2 + \ldots + 
\epsilon^n\,\K_n
\]
and conserved up to an accuracy of $\ob(\epsilon^{n+1})$:
\begin{equation}
\label{math:Rn}
\Rs_n \equiv \K^{(n)}[p',q'] - \K^{(n)}[p,q] = \ob(\epsilon^{n+1}),
\end{equation}
where $\Rs_n$ is referred to as the {\it residual of an approximate
invariant}, or simply the {\it residual}.
To obtain the general solution, we adopt an ansatz where each
$\K_m$ is a polynomial of degree $(m+2)$ with terms of equal powers
in $p$ and $q$:
\[
\begin{array}{l}
\K_0 = C_{2,0}\,p^2 + C_{1,1}\,p\,q\,+ C_{0,2}\,q^2,                    \\[0.25cm]
\K_1 = C_{3,0}\,p^3 + C_{2,1}\,p^2 q + C_{1,2}\,p\,q^2 + C_{0,3}\,q^3,  \\[0.25cm]
\K_2 = C_{4,0}\,p^4 + C_{3,1}\,p^3 q + C_{2,2}\,p^2q^2 + C_{1,3}\,p\,q^3
     + C_{0,4}\,q^4,                                                    \\[0.25cm]
\cdots.
\end{array}
\]
Here, $C_{i,j}$ are coefficients to be determined by satisfying
Eq.~(\ref{math:Rn}).
At first glance, this approach seems to require solving for
$(m+3)$ coefficients for each $\K_m$.
However, by using the following three propositions ---
each straightforward to verify --- we show that the number of
coefficients defined by Eq.~(\ref{math:Rn}) is reduced to
$\lfloor(m+1)/2\rfloor$.

\begin{prop}
\label{prop1}
For the mapping in the form~(\ref{math:Ms}), $\Rs_n$ is of order
$\ob(\epsilon^{n+1})$ only if $C_{i,j} = C_{j,i}$.
\end{prop}
This implies that the first symmetry, $\K[p,q]=\K[q,p]$, is
inherently satisfied, while Eq.~(\ref{math:KK'}) provides a
quantitative measure of the extent to which the second symmetry
is upheld:
\[
\begin{array}{l}
\K^{(n)}[p',q'] -	\K^{(n)}[p,q] =
\K^{(n)}[q',p'] -	\K^{(n)}[p,q] =       \\[0.25cm]
\qquad =\, \K^{(n)}[p,-q+f(p)] -	\K^{(n)}[p,q].
\end{array}
\]
Consequently, we express the expansion as follows for even and odd
orders, respectively, with $m \geq 0$:
\[
\begin{array}{ll}
\K_{2m}   &\!\!\!=
    \sum_{j=0}^{m+1} B_{m+1-j,j}\,\Pi^{m+1-j}\,\cs^j, \\[0.25cm]
\K_{2m+1} &\!\!\!=
    \sum_{j=0}^{m+1} A_{m+1-j,j}\,\Pi^{m+1-j}\,\cs^j\,\Sigma.
\end{array}
\]
\begin{prop}
\label{prop2}
For the mapping in the form~(\ref{math:Ms}), $\Rs_n$ is of order
$\ob(\epsilon^{n+1})$ only if all coefficients of terms
proportional to $\Sigma\,\cs^j$ and $\Pi\,\cs^j$ vanish:
\[
	\forall\,j: \qquad A_{0,j} = 0,\qquad B_{1,j} = 0.
\]
\end{prop}

\begin{prop}
\label{prop3}
If an even polynomial $\K^{(2n)}[p,q]$ satisfies
\[
\K^{(2n)}[p',q'] - \K^{(2n)}[p,q] = \ob(\epsilon^{2n+1}),
\]
for the map~(\ref{math:Ms}), then the modified polynomial
\[
	\K^{(2n)}[p,q] + \epsilon^{2n}\,C_n\,\cs^{n+1}
\]
also satisfies this condition for any value of $C_n$.
\end{prop}

With these propositions in hand, we now look for the solution in
the form:
\[
\begin{array}{l}
	\K_0 = C_0\,\cs                                        \\[0.2cm]
	\K_1 = A_1^1\,\Pi\,\Sigma                              \\[0.2cm]
	\K_2 = B_1^1\,\Pi^2 + C_1\,\cs^2                       \\[0.2cm]
	\K_3 = A_2^2\,\Pi^2\,\Sigma + A_2^1\,\Pi\,\Sigma\,\cs  \\[0.2cm]
	\K_4 = B_2^2\,\Pi^3 + B_2^1\,\Pi^2\,\cs + C_2\,\cs^3   \\[0.2cm]
	\K_5 = A_3^3\,\Pi^3\,\Sigma + A_3^2\,\Pi^2\,\Sigma\,\cs
             + A_3^1\,\Pi\,\Sigma\,\cs^2                       \\[0.2cm]
	\cdots,
\end{array}
\]
where the coefficients are defined using the final notations:
\[
A_i^j \equiv A_{j,i-j},\quad
B_i^j \equiv B_{j+1,i-j},\quad
C_i \equiv B_{0,i+1}.
\]
Thus, at the $n$-th order, we determine the coefficients by
independently setting to zero all combined multipliers of the
terms $p^k q^l\epsilon^{k+l-2}$ in Eq.~(\ref{math:Rn}), where
$k+l-2 \leq n$.
According to the last proposition, the resulting system is
overdetermined, allowing only the coefficients $A_i^j$ and
$B_i^j$ to be uniquely determined.
In contrast, the coefficients $C_i$ cannot be determined without
introducing additional constraints.
Remarkably, Eq.~(\ref{math:Rn}) is satisfied to the appropriate
order regardless of the specific values of $C_i$.

To understand the nature of $C_i$, recall that if $\K[p,q]$ is an
exact invariant of the map $\T$, then any function of $\K[p,q]$
is also an invariant.
For instance, in the case of a linear map $f(p)=a\,p$ with the
invariant
\[
\K_0[p,q]=\cs,
\]
any power series in $\cs$ (whether finite, infinite, or simply
scaled by a constant) remains an invariant:
\[
    C_0\cs + C_1\cs^2 + C_2\cs^3 + \ldots.
\]
Next, consider the symmetric McMillan map~\cite{mcmillan1971problem,IR2001I,IR2002II,IR2003III,
zolkin2024MCdynamics,zolkin2024MCdynamicsIII} with
\[
f_\mathrm{SM}(p) =-\frac{\beta\,p-a}{\alpha\,p^2 + \beta\,p + 1}\,p
\]
and the corresponding invariant
\[
\K_\mathrm{SM}[p,q] = \cs + \beta\,\Pi\,\Sigma + \alpha\,\Pi^2.
\]
Examining a series of the form
\[
C_0\,\K_\mathrm{SM} + C_1\,\K_\mathrm{SM}^2 + C_2\,\K_\mathrm{SM}^3
+ \ldots
\]
and retaining terms up to $\epsilon^2$, we obtain
\[
C_0\,\K_\mathrm{SM} + C_1\cs^2 + \ldots.
\]
It follows that the specific values of $C_i$ do not affect
intrinsic dynamical properties such as the rotation number $\nu_0$
or the twist coefficient $\tau_0 = (\dd\nu/\dd J)_{J=0}$.

\subsection{\label{sec:Lower}Perturbation theory at lower orders}

To begin, we derive exact expressions for the coefficients 
$A_i^j$ and $B_i^j$ in the lower orders of the perturbation theory.
As a first step, we examine the zeroth-order approximate invariant:
\[
	\K^{(0)} = C_0\,\cs,
\]
where  $C_0$ is referred to as the {\it seed coefficient}.
For simplicity, we set the seed to unity, $C_0 = 1$, since all
higher-order terms are proportional to $C_0$.
This choice will later be revisited in the next subsection, which
addresses the removal of singularities.

Using the iterative procedure of setting the multipliers in
Eq.~(\ref{math:Rn}) to zero for $n>0$, we calculate the
approximate invariant up to the third order:
\begin{equation}
\label{math:GS3}
\K^{(3)} = \K_0 + \epsilon\,\K_1 + \epsilon^2\,\K_2 + \epsilon^3\,\K_3
\end{equation}
where the terms are given by:
\[
\begin{array}{l}
\ds \K_1 =-\frac{b}{r_3}\,\Pi\,\Sigma,                              \\[0.35cm]
\ds \K_2 = \frac{b^2-r_3\,c}{r_3\,r_4}\,\Pi^2 + C_1\,\cs^2,         \\[0.35cm]
\ds \K_3 = \frac{\mathcal{T}_0}{r_3\,r_4\,r_5}\,
        \left[
            \Pi^2\,\Sigma - \frac{\Pi\,\Sigma\,\cs}{r_3}
        \right] +
        2\,b\,C_1\frac{\Pi\,\Sigma\,\cs}{r_3},                      \\[0.45cm]
\ds\mathcal{T}_0\,=
    b^3 - (r_3+r_4)\,b\,c + r_3\,r_4\,d.
\end{array}
\]
For completeness, the next-order terms are provided in
Appendix~\ref{secAPP:GSapprox}.
In deriving the expressions above, we introduced the following
notations for resonant terms (see Table~\ref{tab:Rn} for details):
\[
    r_k = \prod_l (a-a_{\nu_l}),        \qquad
    a_{\nu_l} = 2\,\cos(2\,\pi\,\nu_l), \qquad
    \nu_l = l/k,
\]
where the product runs over indices $l$ such that $\nu_l$ are
irreducible fractions with denominator $k$:
\[
    l:\,\mathrm{gcd}(l,k)=1
    \quad\text{and}\quad
    0 \leq \nu_l \leq 1/2.
\]
A dynamical system is said to be on the $k$-th order resonance
when one of the conditions $r_k = 0$, is satisfied.

\subsection{\label{sec:Averaging}Averaging procedure}

After determining all $A_i^j$ and $B_i^j$, we now
discuss the choice of the coefficients $C_i$, using $C_1$  as an
illustrative example.
From the last proposition, we know that at each order, the
residual $\Rs_n$ is of the order $\mathcal{O}(\epsilon^{n+1})$,
regardless of the choice of $C_i$.
This suggests that we should consider the next-order
terms in the residual:
\[
\Rs_n =
    \overline{\Rs_n}\,\epsilon^{n+1} +
    \overline{\overline{\Rs_n}}\,\epsilon^{n+2} +
    \ob(\epsilon^{n+3}).
\]

\begin{table}[t]
\begin{tabular}{p{1cm}p{1.5cm}p{2.5cm}p{3cm}}
\hline\hline
$k$ &$\nu_l$& $a_{\nu_l}$ 		  & $r_k$     \\ \hline
1   & 0     & 2				      & $a-2$     \\
2   & 1/2	&-2				      & $a+2$     \\
3   & 1/3	&-1				      & $a+1$     \\
4   & 1/4	& 0				      & $a$       \\
5   & 1/5	& $(-1+\sqrt{5})/2$   & $a^2+a-1$ \\
    & 2/5	& $(-1-\sqrt{5})/2$   &           \\
6   & 1/6	& 1				      & $a-1$     \\
7   & 1/7	& $2\,\cos(2\,\pi/7)$ &	$a^3+a^2-2\,a-1$  \\
    & 2/7	& $2\,\cos(4\,\pi/7)$ &           \\
    & 3/7	& $2\,\cos(6\,\pi/7)$ &           \\
8   & 1/8	& $\sqrt{2}$          &	$a^2-2$   \\
    & 3/8	& $-\sqrt{2}$         &           \\
9   & 1/9	& $2\,\cos(2\,\pi/9)$ &	$a^3-3\,a+1$      \\
    & 2/9	& $2\,\cos(4\,\pi/9)$ &           \\
    & 4/9	& $2\,\cos(8\,\pi/9)$ &           \\
10  & 1/10	& $(1+\sqrt{5})/2$    &	$a^2-a-1$ \\
    & 3/10	& $(1-\sqrt{5})/2$    &           \\
\hline\hline
\end{tabular}
\caption{
Resonant rotation numbers $\nu_l$,
their corresponding trace values $a_{\nu_l}$, and,
the resonant terms $r_k$ for resonance orders $k \leq 10$.}
\label{tab:Rn}
\end{table}

The coefficient $C_1$ first appears in the second order ($n=2$)
of the approximate invariant Eq.~(\ref{math:GS3}).
The corresponding $\epsilon^3$ terms in the residual $\Rs_2$ are
given by:
\[
\overline{\Rs_2}[p,q] = (a\,p-2\,q)\,p^2\,
\left[ \frac{\mathcal{T}_0}{r_3\,r_4}\,p^2 + 2\,b\,C_1\cs \right].
\]
From this expression, it is evident that no choice of $C_1$ can
entirely eliminate $\overline{\Rs_2}$.
Therefore, instead of seeking to nullify it, we aim to minimize
its variation.

To achieve this, we perform a canonical transformation of the
phase-space variables to Floquet coordinates, defined as:
\[
\begin{array}{ll}
(q,p) \mapsto (J,\phi): & \ds q/\sqrt{2\,J} =
\delta\,\cos\varphi + \frac{a}{2\,\delta}\,\sin\varphi,     \\[0.35cm]
                        & \ds p/\sqrt{2\,J} =
\delta^{-1}\,\sin\varphi,
\end{array}
\]
and where
\[
\delta = \sqrt{\sin(2\,\pi\,\nu_0)} = \left(1-a^2/4\right)^{1/4}.
\]
In these coordinates, the $\cs$ term simplifies to a form
independent of $\varphi$
\[
	\cs = \sqrt{4-a^2}\,J,
\]
while $\overline{\Rs_n}[J,\varphi]$ becomes a periodic function of
$\varphi$.
For example, the expression for $\overline{\Rs_2}[J,\varphi]$ is:
\[
\begin{array}{l}
\ds \overline{\Rs_2}[J,\varphi] =
8\,(-r_1\,r_2)^{1/4}\sin(\varphi)\sin(2\,\varphi)\,J^{5/2}
\times           \\[0.35cm]
\ds	\qquad\times\,
\left[
    \frac{2\,\mathcal{T}_0}{r_1\,r_2\,r_3\,r_4}\sin^2(\varphi) -
    b\,C_1
\right].
\end{array}
\]
Since the phase average of $\overline{\Rs_2}[J,\varphi]$ vanishes:
\[
\int_0^{2\pi} \overline{\Rs_n}[J,\varphi] \,\dd\varphi = 0,
\]

\begin{figure*}[t!]
    \includegraphics[width=\linewidth]{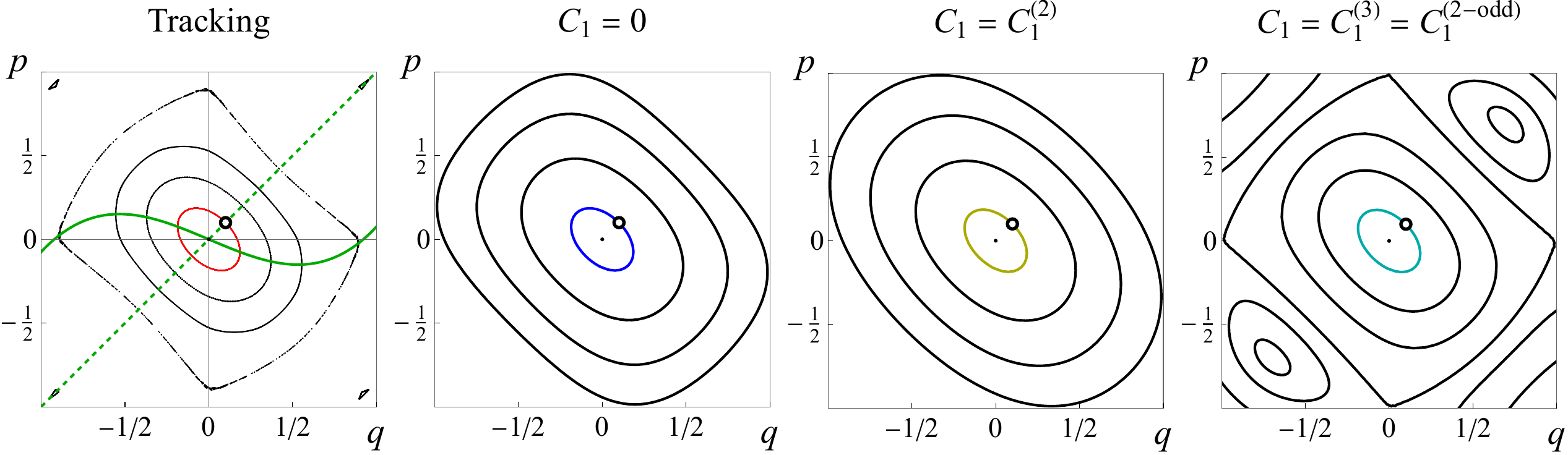}
    \caption{\label{fig:AveragingPS}
    The left plot depicts phase-space trajectories of the cubic
    map $f(p) = a\,p + p^3$ for $a=-17/20$, obtained through
    numerical tracking.
    A sample orbit $\z_0$, corresponding to the initial condition
    $(q_0,p_0)=(1/10,1/10)$ (marked as a white point), is
    highlighted in red.
    Green curves are the first $p=q$ (dashed) and second $p=f(q)/2$
    (solid) symmetry lines.
    The subsequent three plots illustrate the constant level sets
    of the approximate invariant $\K^{(2)}[p,q]$ for different
    values of the coefficient $C_1$.
    The case $C_1 = 0$ represents an integrable approximation via
    the canonical McMillan map.
    The next case with $C_1 = C_1^{(2)}$ demonstrates an
    ``incorrect'' averaging at order $\epsilon^3$.
    The last plot corresponds to the averaging at order
    $\epsilon^4$, yielding the invariant
    $\langle\K^{(2)}_\mathrm{odd}[p,q]\rangle$.
    In each plot, the level set passing through the point $\z_0$
    is highlighted for clarity.
    }
\end{figure*}

\noindent
we minimize the integral of its square:
\[
I_n = \int_0^{2\pi} \overline{\Rs_n}^2[J,\varphi] \,\dd\varphi
\]
by solving for the coefficients $C_k$ using the system of linear
equations:
\[
    \frac{\dd}{\dd C_k}\,I_n = 0.
\]
For the second order ($n=2$), this procedure yields:
\begin{equation}
\label{math:C12GS}
C_1^{(2)} = \frac{5}{4\,b}\,\frac{\mathcal{T}_0}{r_1\,r_2\,r_3\,r_4}.
\end{equation}

This {\it averaging procedure} is repeated at each order of the
perturbation theory.
The superscript notation indicates the order $n$ for which the
coefficient is determined.
For example, at $n=3$, only $C_1^{(3)}$ is involved, and it is
computed by solving $\dd I_3/\dd C_1 = 0$.
At $n=4$, the system of two linear equations
$\dd I_4/\dd C_{1,2}=0$ provides the values of $C_{1,2}^{(4)}$.
Exact expressions for higher-order coefficients are provided in
Appendix~\ref{secAPP:GSav}.

When the force is an odd function, $f(-p)=-f(p)$:
\[
    f_\mathrm{odd}(p) = a\,p + c\,p^3 + e\,p^5 + g\,p^7 + \ldots,
\]
the expansion of $\K^{(n)}$ involves only even powers of $\epsilon$,
yielding the following expression for the first two nonlinear
orders:
\[
\begin{array}{l}
\ds \K^{(n)}_\mathrm{odd}[p,q] =
\K_0 + \epsilon^2\,\K_2 + \epsilon^4\K_4 + \ldots =     \\[0.35cm]
\ds \quad
=\cs-\frac{c}{r_4}\,\Pi^2\epsilon^2 +
C_1\,\cs^2\epsilon^2  +
C_2\,\cs^3\epsilon^4\,+
\\[0.35cm]
\ds \quad\,\, +
\left[
    \frac{c^2-r_4 e}{r_3\,r_4\,r_6}
    \left(
        \Pi - \frac{\cs}{r_4}
    \right) -
    2\,c\,C_1\frac{\cs}{r_4}
\right]\Pi^2\epsilon^4 + \ldots.
\end{array}
\]
In this case, all $\overline{\Rs_n} = 0$, so the leading-order contributions to the residual come from $\epsilon^{n+2}$ terms,
specifically $\overline{\overline{\Rs_n}}$.
For $n=2$, we have
\[
\begin{array}{l}
\ds \overline{\overline{\Rs_2}}[J,\varphi] =
16\,\sin^2(\varphi)\sin(2\,\varphi)\,J^{3}
\times           \\[0.35cm]
\ds	\qquad\times\,
\left[
    2\,\frac{r_4\,e-c^2}{r_1\,r_2\,r_4}\sin^2(\varphi) -
    c\,C_1
\right],
\end{array}
\]
which allows us to obtain
\begin{equation}
\label{math:C12odd}
C_1^{(2\mathrm{-odd})} =
    \frac{7}{5\,c}\,\frac{r_4\,e-c^2}{r_1\,r_2\,r_4}:
\qquad\qquad
\frac{\dd I_2^\mathrm{odd}}{\dd C_1} = 0,
\end{equation}
where
\[
I_n^\mathrm{odd} = \int_0^{2\pi}
    \overline{\overline{\Rs_n}}^{\,2}[J,\varphi] \,\dd\varphi.
\]
Notice that, since the averaging was performed over $\epsilon^4$
terms, Eq.~(\ref{math:C12odd}) cannot be obtained as a limiting
case $b,d=0$ of the general expansion $C_1^{(2)}$,
Eq.~(\ref{math:C12GS}).
Instead, it arises from applying the same limit to $C_1^{(3)}$,
see Appendix~\ref{secAPP:GSav}.

\begin{figure}[t!]
    \includegraphics[width=\linewidth]{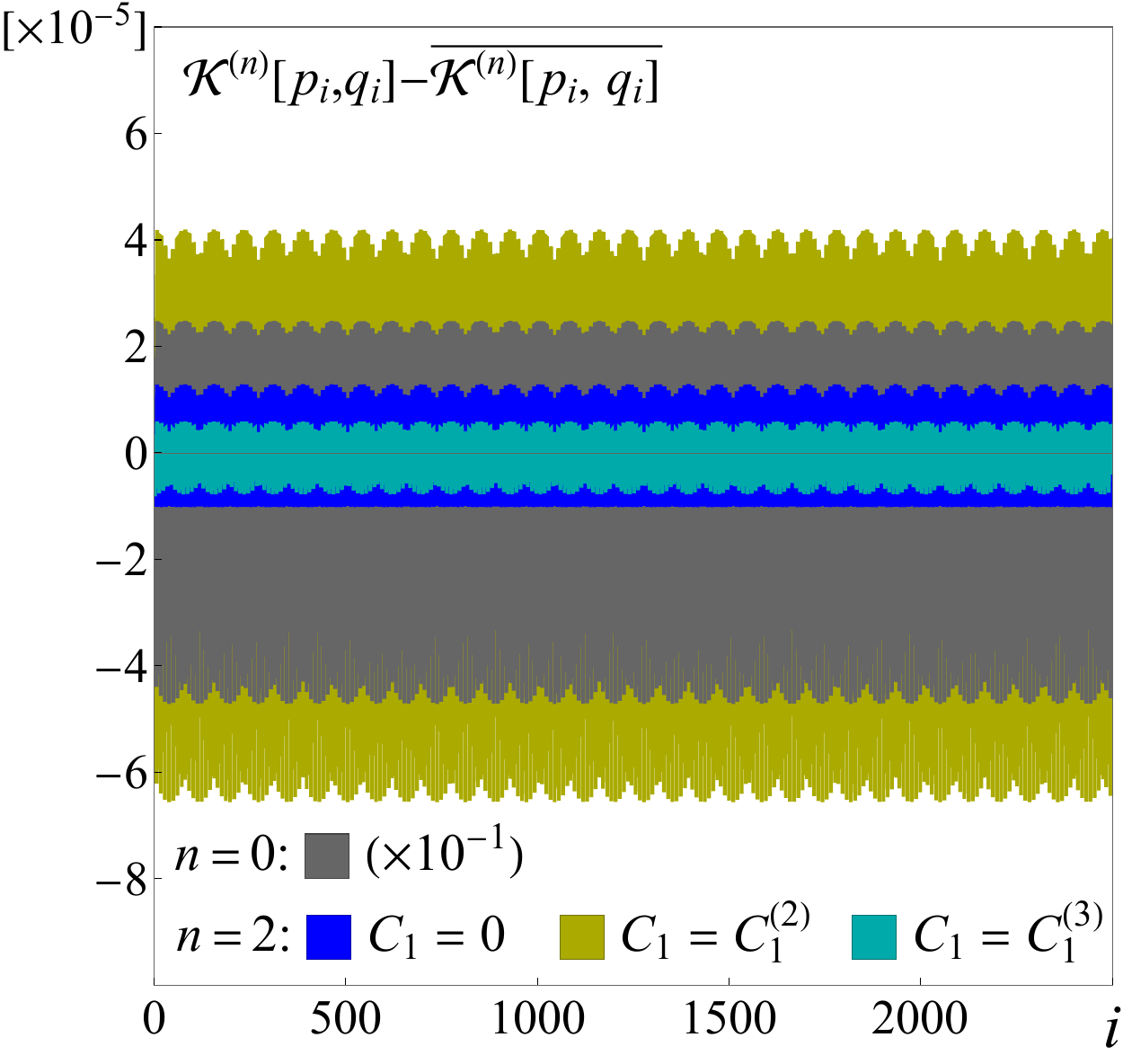}\vspace{-0.35cm}
    \caption{\label{fig:AveragingK2}
    The approximate invariant $\K^{(n)}[p_i,q_i]$, evaluated
    along the sample orbit $(q_0,p_0)=(1/10,1/10)$ for the cubic
    map $f(p) = a\,p + p^3$ with $a=-17/20$, is shown as a
    function of the number of iterations $i$.
    The vertical axis is scaled by a factor of $\times 10^{-5}$
    for clarity.
    Different colors represent $n=0$ and $n=2$ for various choices
    of the coefficient $C_1$, as detailed in
    Fig.~\ref{fig:AveragingPS}.
    For each case, the average value of the evaluated invariant is subtracted to facilitate direct comparison.
    }
\end{figure}

\vspace{0.1cm}
Figs.~\ref{fig:AveragingPS} and \ref{fig:AveragingK2} illustrate
the impact of different choices of $C_1$ on the cubic H\'enon map,
given by $f=a\,p+p^3$ with $a=-17/20$.
The first set of figures compares the trajectory of a sample orbit,
initialized at $\z_0=10^{-1}(1,1)$ (shown in red in the left plot),
with the level sets of the approximate invariant of second order
$\K^{(2)}[\z_0]$.
The second plot in Fig.~\ref{fig:AveragingPS} corresponds to
$C_1 = 0$, representing the approximation via the integrable
McMillan map.
The third graph illustrates the effect of an ``incorrect'' averaging,
while the fourth shows the case for $C_1 = C_1^{(2-\mathrm{odd})}$.
Fig.~\ref{fig:AveragingK2} further demonstrates the conservation of
the approximate invariant along the sample orbit, with the average
value subtracted to allow for a clearer comparison.
As observed, the choice $C_1 = C_1^{(3)} = C_1^{(2-\mathrm{odd})}$
provides the most accurate results.

\newpage
After determining the coefficients $C_i$, we define, at each order,
an {\it averaged approximate invariant}:
\[
\langle \K^{(n)} \rangle = \K^{(n)}(C_k:\dd I_n/\dd C_k = 0).
\]
For $n=2$, this results in:
\[
\begin{array}{l}
\ds	\langle \K^{(2)}\rangle[p,q]\,=
	\cs -
	\frac{b}{r_3}\,\Pi\,\Sigma\,\epsilon\,+         \\[0.35cm]
\ds \qquad\qquad\quad\, + \left[
    \frac{b^2 - r_3\,c}{r_3\,r_4}\,\Pi^2 +
    \frac{5}{4\,b}\,\frac{\mathcal{T}_0}{r_1\,r_2\,r_3\,r_4}\,\cs^2
    \right]\epsilon^2,                                  \\[0.35cm]
\ds	\langle \K^{(2)}_\mathrm{odd}\rangle[p,q] =
\cs -  \left[
    \frac{c}{r_4}\,\Pi^2 +
    \frac{7}{5\,c}\,\frac{c^2-r_4\,e}{r_1\,r_2\,r_4}\,\cs^2
\right]\epsilon^2.
\end{array}
\]
Before moving to the next subsection, we present two observations
worth mentioning.

\vspace{0.1cm}\noindent
{\bf Observation \#1}.
Odd mappings in the McMillan form exhibit {\it double reversibility}
because they commute with the area-preserving involution
\cite{roberts1992revers,zolkinHenonSet}:
\[
\T_\mathrm{odd} = \Rt(\pi)\circ\T_\mathrm{odd}\circ\Rt^{-1}(\pi).
\]
This property gives rise to a class of transformations distinct
from $\R_{1,2}$, such that:
\[
    \T_\mathrm{odd} = \mathrm{Q}_2 \circ \mathrm{Q}_1,
    \qquad
    \T_\mathrm{odd}^{-1} = \mathrm{Q}_1 \circ \mathrm{Q}_2,
    \qquad
    \mathrm{Q}_{1,2}^2 = \mathrm{Id},
    \]
where
$\mathrm{Q}_1 = \R_1\circ\Rt(\pi) = \Rt(3\pi/4)$ and
$\mathrm{Q}_2 = \T_\mathrm{odd}\circ\mathrm{Q}_1$,
\[
\begin{array}{ll}
\mathrm{Q}_1:   & q' =-p,                   \\[0.25cm]
                & p' =-q,
\end{array}
\qquad\qquad
\begin{array}{ll}
\mathrm{Q}_2:   & q' =-q,                   \\[0.25cm]
                & p' = p - f_\mathrm{odd}(q).
\end{array}
\]
The transformations $\mathrm{Q}_{1,2}$ form an independent
group of symmetries, introducing two additional symmetry
lines:
\[
l_3 = \mathrm{Fix}\,(\mathrm{Q}_1):\,\,
    p =-q,                  \qquad
l_4 = \mathrm{Fix}\,(\mathrm{Q}_2):\,\,
    q = 0.
\]
It is worth noting that the absence of odd powers of $\epsilon$
in $\K^{(n)}_\mathrm{odd}$ enforces symmetry along $l_3$ at each
order.
Thus, double reversibility offers an additional constraint that
can be utilized in determining the approximate invariant.

\vspace{0.1cm}\noindent
{\bf Observation \#2}. 
As a result of choosing $C_0 = 1$, both $\K^{(n)}$ and
$\langle \K^{(n)}\rangle$ contain singular terms with denominators
that vanish at certain resonant values of $a_{\nu_j}$: $r_j = 0$.
Before averaging, some terms in $\K^{(n)}$ involve denominators
proportional to $r_j^m$ where $m>1$.
However, after applying the averaging procedure and substituting
all $C_i$, the powers of these terms appear to reduce to $m=1$ in
all orders.
Consequently, $\langle \K^{(n)}\rangle$ can be expressed as:
\[
\langle \K^{(n)} \rangle = 
    \frac{1}{r_1\,r_2 \ldots r_{n+2}}\,\Big( \ldots \Big),
\]
where the terms inside the parentheses do not introduce
singularities at rational values of $\nu_0$.

\subsection{\label{sec:Removal}Removal of singularities}

Since $r_j$ are independent of the dynamical variables, we can
choose $C_0$ to be the product of all singular denominators that
appear in the expressions.
Thus, we define {\it non-singular} averaged invariants:
\[
\{\K^{(n)}\} = \langle \K^{(n)} \rangle\prod_{k=1}^{n+2} r_k.
\]
For instance, compare the following with the previously derived
results for $n=2$:
\[
\begin{array}{l}
\ds	\{\K^{(2)}\}[p,q]\,=
	r_1\,r_2\,r_3\,r_4\,\cs -
	r_1\,r_2\,r_4\,b\,\Pi\,\Sigma\,\epsilon\,+     \\[0.3cm]
\ds	\qquad+\left[
		r_1\,r_2\,(b^2 - r_3\,c)\,\Pi^2 +
		\frac{5}{4}\,\frac{\mathcal{T}_0}{b}\,\cs^2
	\right]\epsilon^2,                             \\[0.4cm]
\ds	\{\K^{(2)}_\mathrm{odd}\}[p,q] =
	r_1\,r_2\,r_4\,\cs\,-                          \\[0.3cm]
\ds	\qquad-\left[
		r_1\,r_2\,c\,\Pi^2 +
		\frac{7}{5}\,\frac{c^2-r_4\,e}{c}\,\cs^2
	\right]\epsilon^2.
\end{array}
\]
In this new form, when the resonance condition $r_j = 0$ is met,
the previously singular terms now define the shape of the invariant,
while all other terms effectively vanish due to the corresponding
multiplier $r_j$.

\vspace{0.1cm}\noindent
{\bf Observation \#3}. As we proceed to higher orders, it becomes
evident that, right at resonance $r_j = 0$, lower-order terms
proportional to $\cs^m$, including $m=1$, can vanish.
This is an important property that perturbation theory should
exhibit, as in nonlinear systems, a zeroth-order linearized
invariant $\cs$ may be an inadequate approximation, even for
infinitesimally small amplitudes.

\begin{figure*}[t!]
    \includegraphics[width=\linewidth]{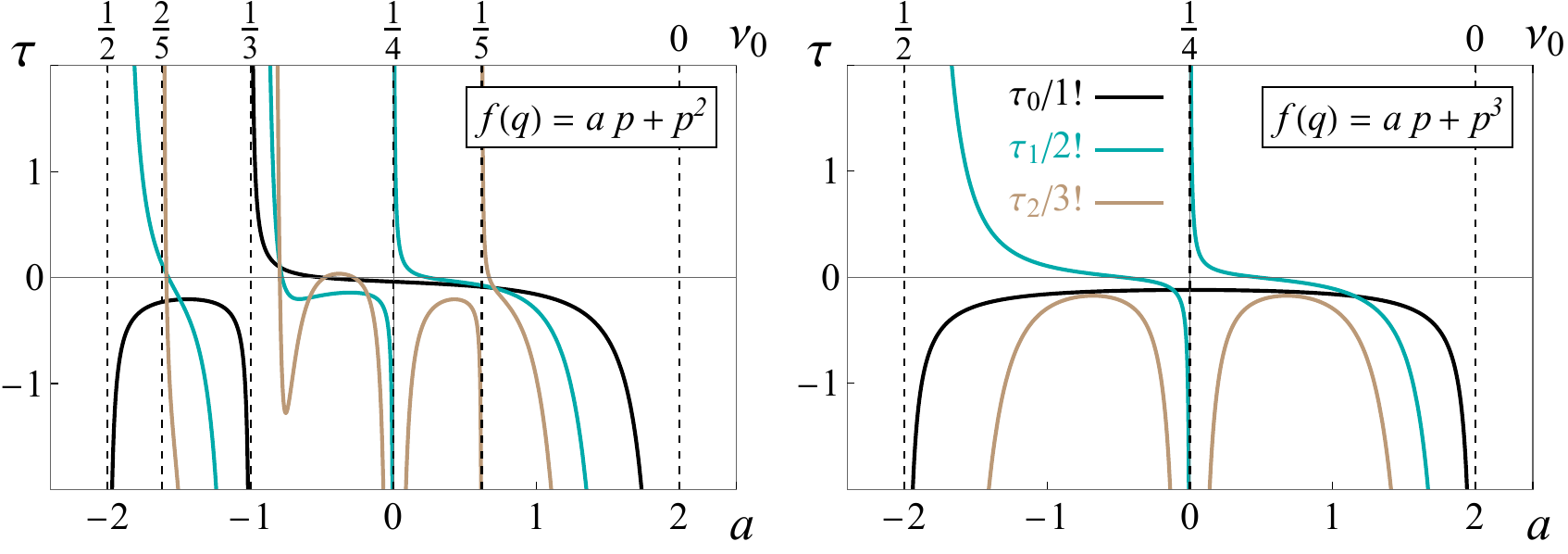}\vspace{-0.3cm}
    \caption{\label{fig:Twists}
    First three twist coefficients $\tau_{0,1,2}$ for the H\'enon
    quadratic ($f(p) = a\,p + p^2$) and cubic ($f(p) = a\,p + p^3$)
    mappings.
    The linear scale at the bottom provides the trace of the
    Jacobian at the origin, $a$, while the scale at the top shows
    corresponding values of the bare rotation number, $\nu_0$,
    defined for $|a| \leq 2$.
    }\vspace{-0.45cm}
\end{figure*}

Before presenting examples, let us recall that in Birkhoff normal
form theory (see, e.g., \cite{dullin2000twistless}) the map’s
{\it canonical form} is given by
\begin{equation}
\label{math:JTheta}    
\begin{array}{l}
    J' = J,                 \\[0.25cm]
    \theta' = \theta + 2\,\pi\,\nu(J),
\end{array}
\end{equation}
where $J$ is the {\it action} (or {\it symplectic radius}) and
$\theta$ is its {\it conjugate angle} variables.
The rotation number is often expanded as a power series in the
action:
\[
\nu(J) = \nu_0 + \tau\,J
    = \nu_0 + \tau_0 J + \frac{1}{2!}\,\tau_1 J^2
        + \frac{1}{3!}\,\tau_2 J^3 + \mathcal{O}(J^4),
\]
where its derivative, known as the {\it twist}, is given by
\[
\tau(J) = \frac{\dd \nu}{\dd J} = \tau_0 + \tau_1 J +
        \frac{1}{2}\,\tau_2 J^2
        + \mathcal{O}(J^3).
\]
The coefficients $\tau_i$ are called {\it twist coefficients}.
For McMillan-form mappings~(\ref{math:Ms}) with a smooth force,
the first twist coefficient $\tau_0$ is expressed as:
\begin{equation}
\label{math:twist}
2\,\pi\,\tau_0 = \frac{1}{r_1\,r_2}\,\left[
    3\,c - 4\,\frac{a+1/2}{r_1\,r_3}\,b^2
\right].
\end{equation}
When $b \neq 0$, $\tau_0$ is well-defined except at
$\nu_0 = 0,\,1/2,\,1/3$ (i.e., $a = 2,-2,-1$), where it becomes
singular.
Additionally, $\tau_{1,2}$ is only defined for
$\nu_0\neq 1/4,1/5,2/5$ (i.e., $a\neq 0,(-1\pm\sqrt{5})/2$).
Expressions for $\tau_1$ and $\tau_2$ are provided at the end of
this article, Eqs.~(\ref{math:Twist12}) and ~(\ref{math:Twist3}),
while Fig.~\ref{fig:Twists} illustrates $\tau_{0,1,2}(a)$ for the
H\'enon quadratic and cubic maps, which we will use in our
examples.
While these results can be derived using various methods --- such as
Birkhoff normal forms~\cite{dullin2000twistless},
Lie algebra techniques~\cite{bengtsson1997, morozov2017dynamical},
square matrix method~\cite{hua2017square}, or
Deprit perturbation theory~\cite{michelotti1995intermediate} ---
recent studies~\cite{zolkin2024MCdynamics,zolkin2024MCdynamicsIII}
have demonstrated that $\tau_0$ can also be obtained from expansions
of the action variable in integrable McMillan multipoles.
This provides additional qualitative insights and serves as a useful
tool for further discussions.

\subsubsection{Quarter integer resonance}

\begin{figure}[t!]
    \includegraphics[width=\linewidth]{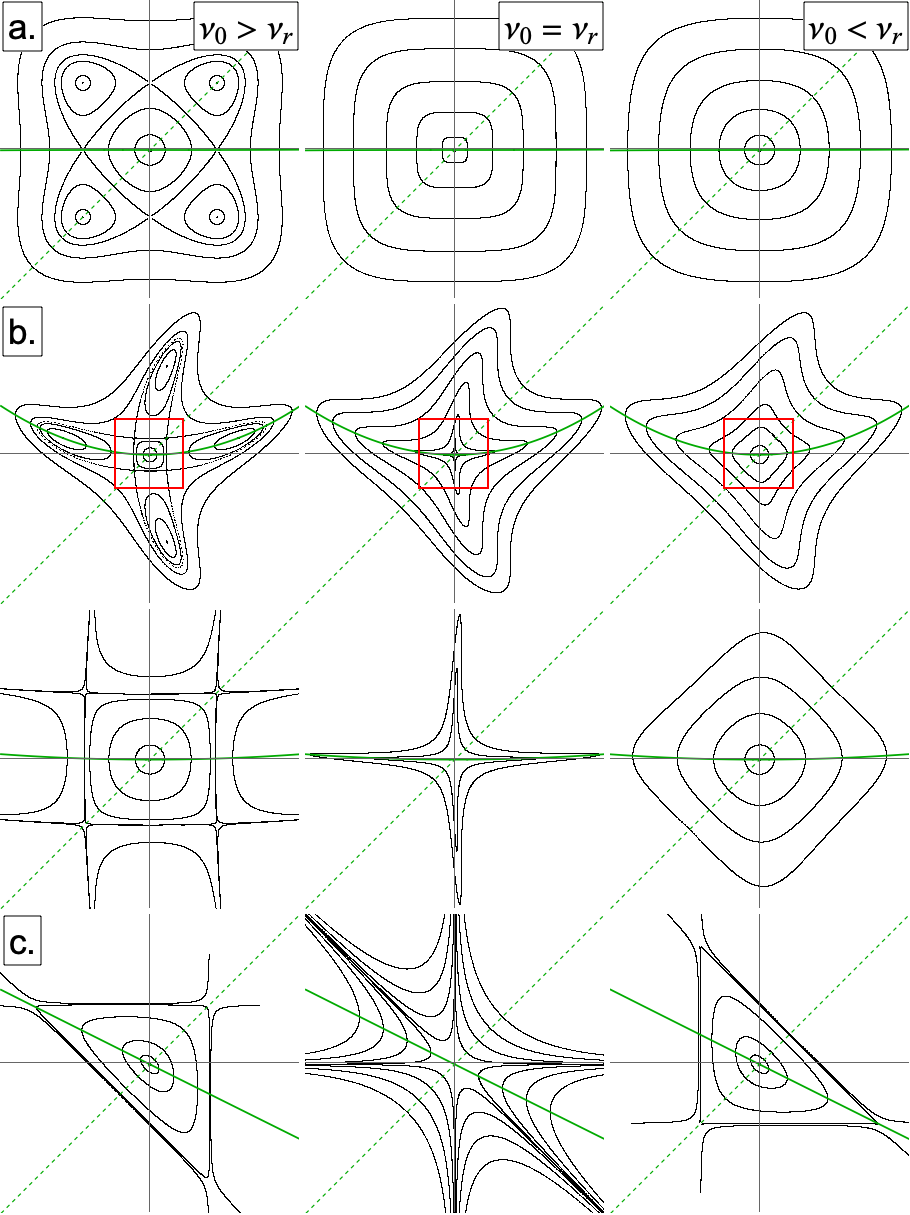}
    \caption{\label{fig:RESs}
    Phase space morphology across low-order resonances.
    The top rows show the 4-island chain bifurcation $\nu_r=1/4$
    for the H\'enon (a.) cubic $f(p) = a\,p+p^2$ and
    (b.) quadratic $f(p) = a\,p+p^3$ maps.
    The third row provides a magnified view of the regions
    highlighted in red.
    The last row (c.) depicts the touch-and-go bifurcation in
    the quadratic map, $\nu_r = 1/3$.
    Green curves are the first (dashed) and second (solid)
    symmetry lines.
    }\vspace{-0.2cm}
\end{figure}

As our first example, we examine the quarter-integer resonance
$\nu_0 = 1/4$ at $a = 0$.
For both the quadratic and cubic maps, the first twist coefficient
remains finite:
\[
    \tau_0[b=1,c=0] = -\frac{1}{8\,\pi},
    \qquad
    \tau_0[b=0,c=1] = -\frac{3}{8\,\pi}.
\]
This ensures that the nonlinear resonance is stable, leading to
the formation of a characteristic chain of four islands for
$\nu_0 > \nu_r$, as shown in cases (a.) and (b.) of
Fig.~\ref{fig:RESs}.
Meanwhile, in the general case $b^2 \neq c$, higher-order twist
coefficients become singular.

In the vicinity of the resonance, $\delta_r=\nu_r-\nu_0\approx 0$,
one might naively expect small-amplitude invariant curves described
by
\[
\cs[a=0] = p^2 + q^2.
\]
However, a closer inspection reveals a more intricate picture.
For the H\'enon quadratic map at exact resonance $\nu_0 = 1/4$,
phase space trajectories exhibit cross-like structures, as seen
in Fig.~\ref{fig:RESs} (b.).
In contrast, zooming in on the region near the origin for
slightly off-resonant cases ($\delta_r \neq 0$) reveals circular
trajectories, consistent with expectations.
A similar phenomenon occurs in the cubic map: when on the resonance,
the trajectories form more square-like invariant structures rather
than circular ones, as shown in Fig.~\ref{fig:RESs} (a.).
To better understand these emergent shapes, we turn to perturbation
theory and examine the first few orders of the expansion.
After averaging and eliminating singular terms, we obtain the
following expressions for the case $c=1$, up to a constant
multiplier:
\[
\begin{array}{l}
\ds \{\K_{c=1}^{(2)}\} = \left(
    \Pi^2 - \frac{7}{10}\frac{\cs^2}{2}
\right)\epsilon^2,                                  \\[0.3cm]
\ds \{\K_{c=1}^{(4)}\} = \left(
    \Pi^2 - \frac{\cs^2}{2}
\right)\epsilon^2 + \Pi^3\epsilon^4.
\end{array}
\]

\begin{figure}[t!]
    \includegraphics[width=\linewidth]{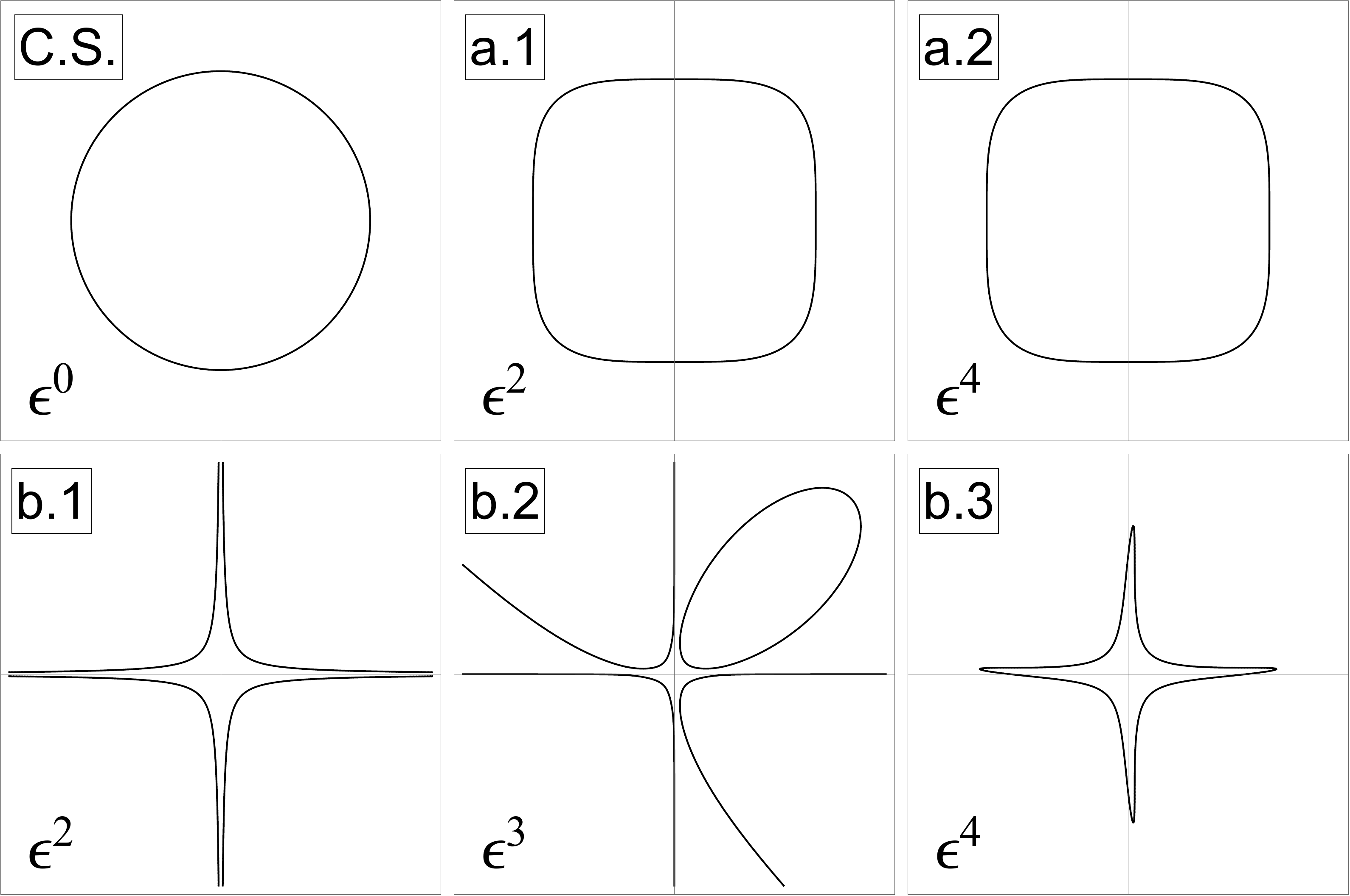}
    \caption{\label{fig:CSr4}
    Invariant curves for small amplitudes
    $\{\K^{(n)}\}[p,q] \approx 0$ at the $r_4 = 0$ resonance for
    the zeroth-order linearized approximation ($\cs$), cubic (a.)
    and quadratic (b.) H\'enon maps.
    Cases (a.1--2) and (b.1--3) illustrate the level sets obtained
    by retaining lower-order terms from the converged expressions
    in Eqs.~(\ref{math:CSr4OC}) and (\ref{math:CSr4SX}), truncated
    up to $\epsilon^n$, as indicated at the bottom left corner of
    each plot.
    }
\end{figure}

As we proceed, we observe that the coefficients provided by
$\{\K_{c=1}^{(4)}\}$ converge:
\begin{equation}
\label{math:CSr4OC}    
\{\K_{c=1}^{(n \geq 6)}\} = \left(
    \Pi^2 - \frac{\cs^2}{2}
\right)\epsilon^2 + \Pi^3\epsilon^4 + \ob(\epsilon^6).
\end{equation}
The first term ($\propto \epsilon^2$) now matches precisely the trajectories for infinitesimally small amplitudes.
The corresponding shape is shown in Fig.~\ref{fig:CSr4}, (a.1).
Since this leading-order term already defines a stable (closed)
level set, the addition of the next-order term, proportional to
$\epsilon^3$, does not alter the shape in the limit
$q,p \rightarrow 0$, case (a.2).

Returning to the case of the quadratic map, we obtain the
following sequence of approximations:
\[
\begin{array}{l}
\ds \{\K_{b=1}^{(1)}\} = \cs - \Pi\,\Sigma\,\epsilon,       \\[0.3cm]
\ds \{\K_{b=1}^{(2)}\} = \left(
    \Pi^2 - \frac{5}{8}\frac{\cs^2}{2}
\right)\epsilon^2,                                  \\[0.3cm]
\ds \{\K_{b=1}^{(3)}\} = \left(
    \Pi^2 - \frac{34}{69}\frac{\cs^2}{2}
\right)\epsilon^2 + \left(
    \Pi - \frac{35}{69}\cs
\right)\Pi\,\Sigma\,\epsilon^3
\end{array}
\]
\begin{equation}
\label{math:CSr4SX}
\begin{array}{l}
\ds \{\K_{b=1}^{(4)}\} =
    \Pi^2\epsilon^2 +
    (\Pi-\cs)\,\Pi\,\Sigma\,\epsilon^3 +
    2\,\Pi^3\epsilon^4,                             \\[0.35cm]
\ds \{\K_{b=1}^{(5)}\} =
    \Pi^2\epsilon^2 +
    (\Pi-\cs)\,\Pi\,\Sigma\,\epsilon^3\,+           \\[0.15cm]
\ds\quad+\,\left(
        2\,\Pi^3-\Pi^2\,\cs + \frac{\cs^3}{3}
    \right)\epsilon^4 +
    (\Pi-\cs)\,\Pi^2\Sigma\,\epsilon^5,             \\[0.2cm]
\end{array}
\end{equation}
such that
$\{\K_{b=1}^{(n \geq 6)}\} = \{\K_{b=1}^{(5)}\}+\ob(\epsilon^6)$,
and where the expansion must be carried out up to $n=5$ to obtain
converged terms that define a closed level set of the invariant.

Fig.~\ref{fig:CSr4}, cases (b.1) -- (b.3), illustrates the level
sets obtained by retaining lower-order terms in
Eq.~(\ref{math:CSr4SX}).
The closer one examines trajectories near the origin, the better
the curve (b.3) aligns with numerical tracking results.
However, in the infinitesimally small amplitude limit, the dominant
term $\Pi^2\epsilon^2$ alone determines the structure, forming
branches of a quadratic hyperbola.

Analyzing the converged terms for a general force function
$f(p) = a\,p + b\,p^2 + c\,p^3 + \ldots$ at the $1/4$ resonance,
we obtain
\[
\{\K_{a=0}\} =
\left[
    (b^2-c)\,\Pi^2 + c\,\frac{\cs^2}{2}
\right]\epsilon^2 + \ldots.
\]
This expression reveals that a stable level set in terms
$\propto \epsilon^2$ does not form only when $c=0$.
Moreover, only when $b^2 = c$, does the small-amplitude limit
correspond to a circular level set, with the expansion beginning
with a $\cs^2$ term rather than $\cs$.
In this case, the higher-order twist coefficients remain finite.
Additional details on the higher-order terms that converge
for this and other low-order resonances can be found in
Appendix~\ref{secAPP:GSres}.

\begin{figure*}[ht!]
    \includegraphics[width=\linewidth]{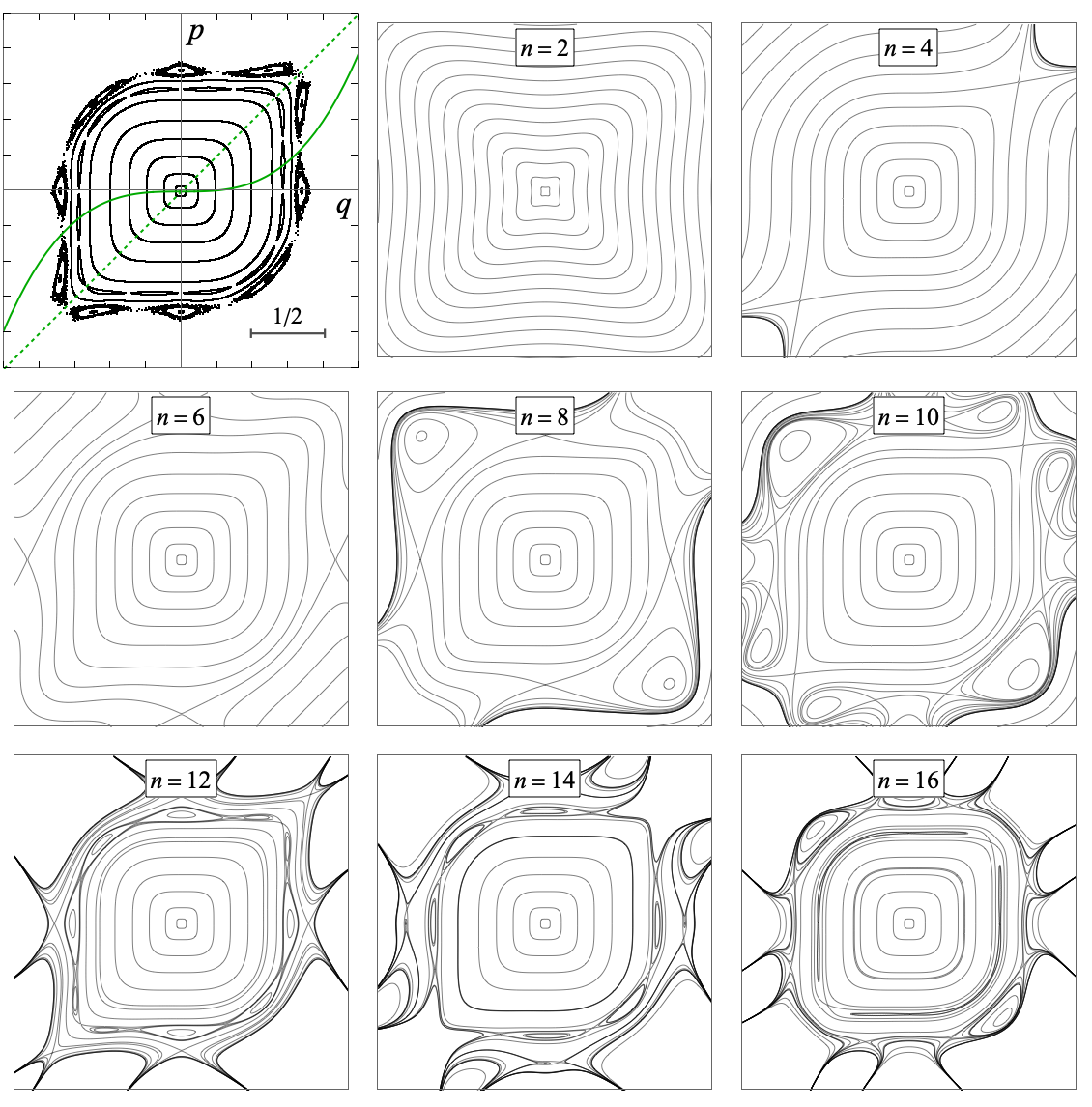}
    \caption{\label{fig:OCTr4}
    Trajectories of the H\'enon cubic map $f(p) = a\,p+p^3$ at
    the $1/4$ integer resonance (top left), compared with
    approximate non-singular averaged invariants
    $\{\K^{(n)}_{c=1}\}[p,q]$.
    The bar scale provides a reference for size, with all
    approximate invariants displayed on the same grid.
    Green curves indicate the first (dashed) and second (solid)
    symmetry lines.
    }
\end{figure*}

Fig.~\ref{fig:OCTr4} compares the results of tracking (top left
plot) with averaged non-resonant approximate invariants computed
up to order $n=16$.
Unlike Fig.~\ref{fig:CSr4}, this figure displays the complete
invariant at each order, rather than only the converged terms.
As $n$ increases, the approximate invariant progressively reveals
finer details of the dominant 10-island structure surrounding the
central stability region.
By $n=12$, all 10 islands become visible.
However, at higher orders, perturbation theory shifts focus to
the next 14-island chain nested within the 10-island structure.
At $n=16$, magnification reveals 10 out of the 14 islands along
with associated saddles.
As a consequence of this infinite sequence of shifts to finer
chains of islands with higher periods, the terms associated with
these resonances fail to converge.

\begin{figure*}[ht]
    \includegraphics[width=0.84\linewidth]{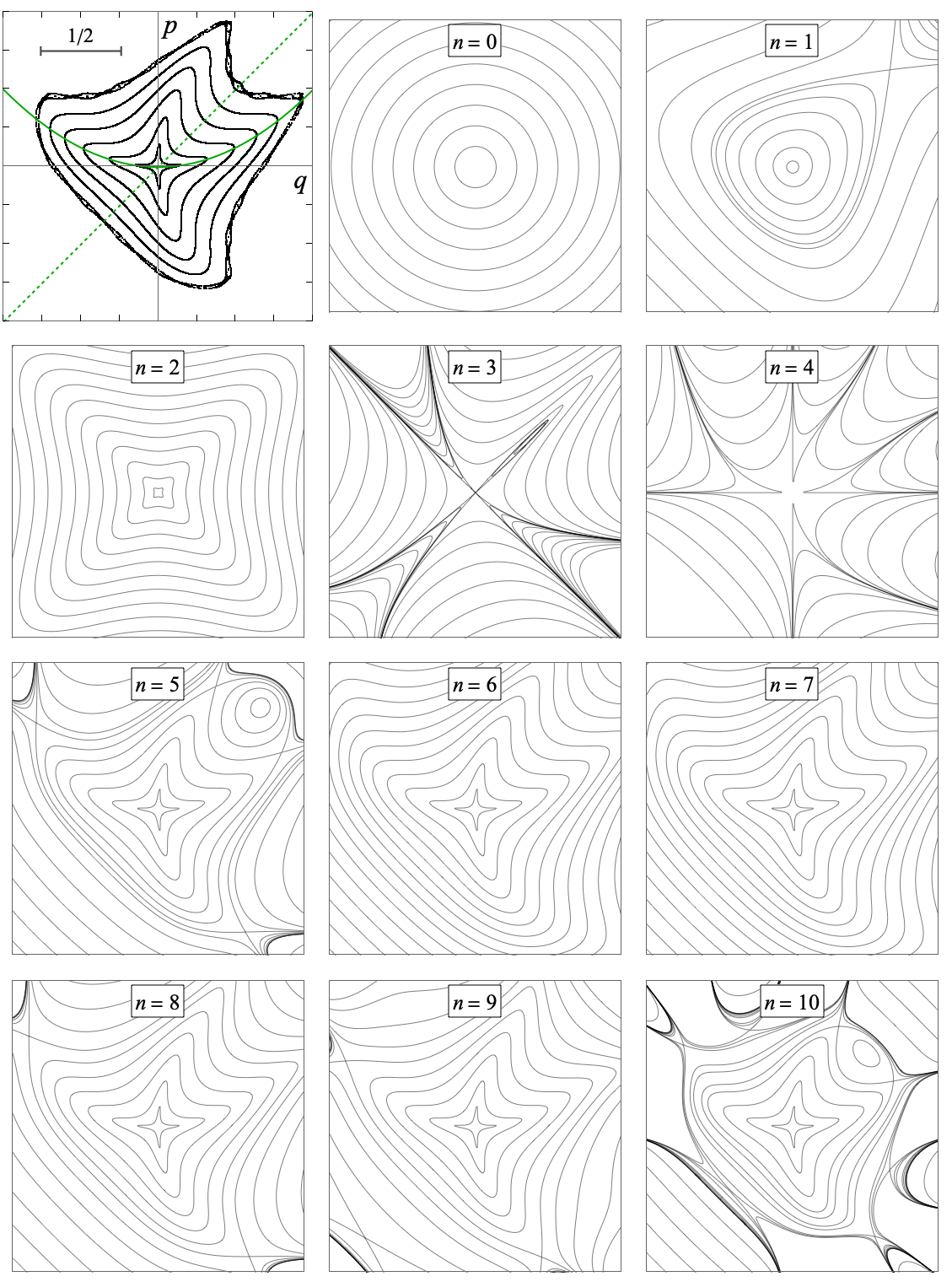}
    \caption{\label{fig:SXTr4}
    Same as Fig.~\ref{fig:OCTr4}, but for the quadratic map
    $f=a\,p+p^2$ at the quarter-integer resonance ($a=0$).
    }
\end{figure*}

Fig.~\ref{fig:SXTr4}, similar to \ref{fig:OCTr4}, presents
averaged non-resonant invariants for quadratic map $f(p) = p^2$.
While the largest

\newpage
\noindent
visible chain on this scale consists of 21
islands, we do not
expect lower-order perturbation theory to capture its full
structure. 
However, for $n=5$ and $n \geq 8$, the simply connected region
surrounding the origin provides a reasonable approximation of the
stable trajectories observed in numerical simulation.
In the final section, we explore this question in greater detail,
extending our analysis beyond the resonant case to the more general
scenario.

\begin{figure*}[th!]
    \includegraphics[width=\linewidth]{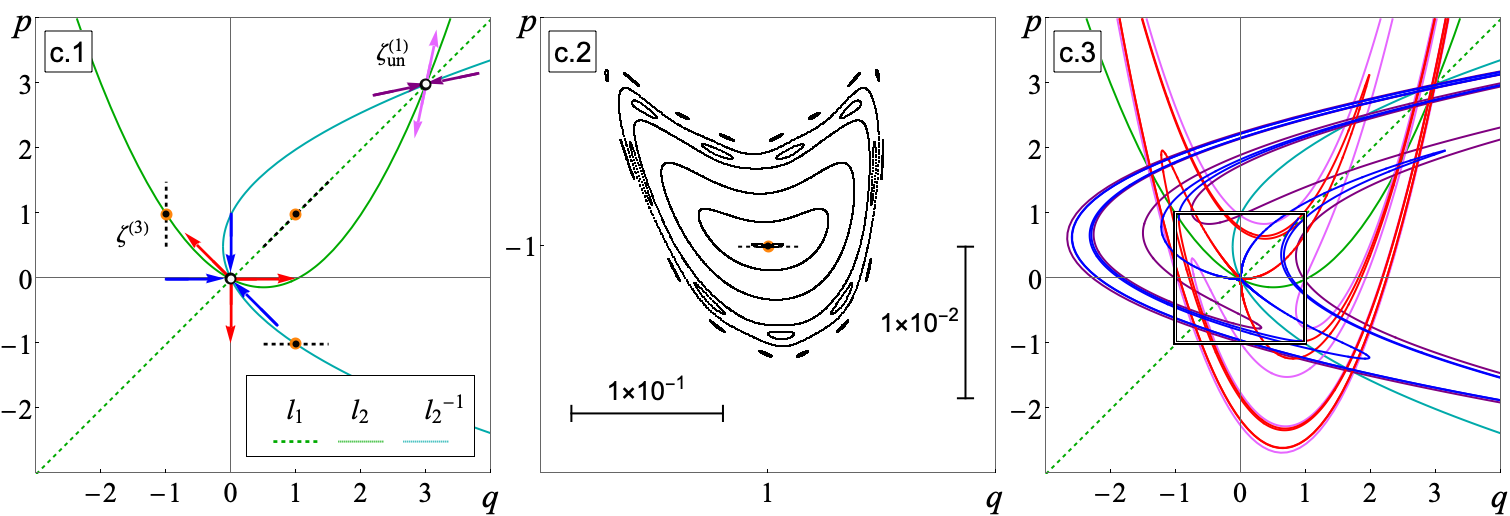}
    \includegraphics[width=\linewidth]{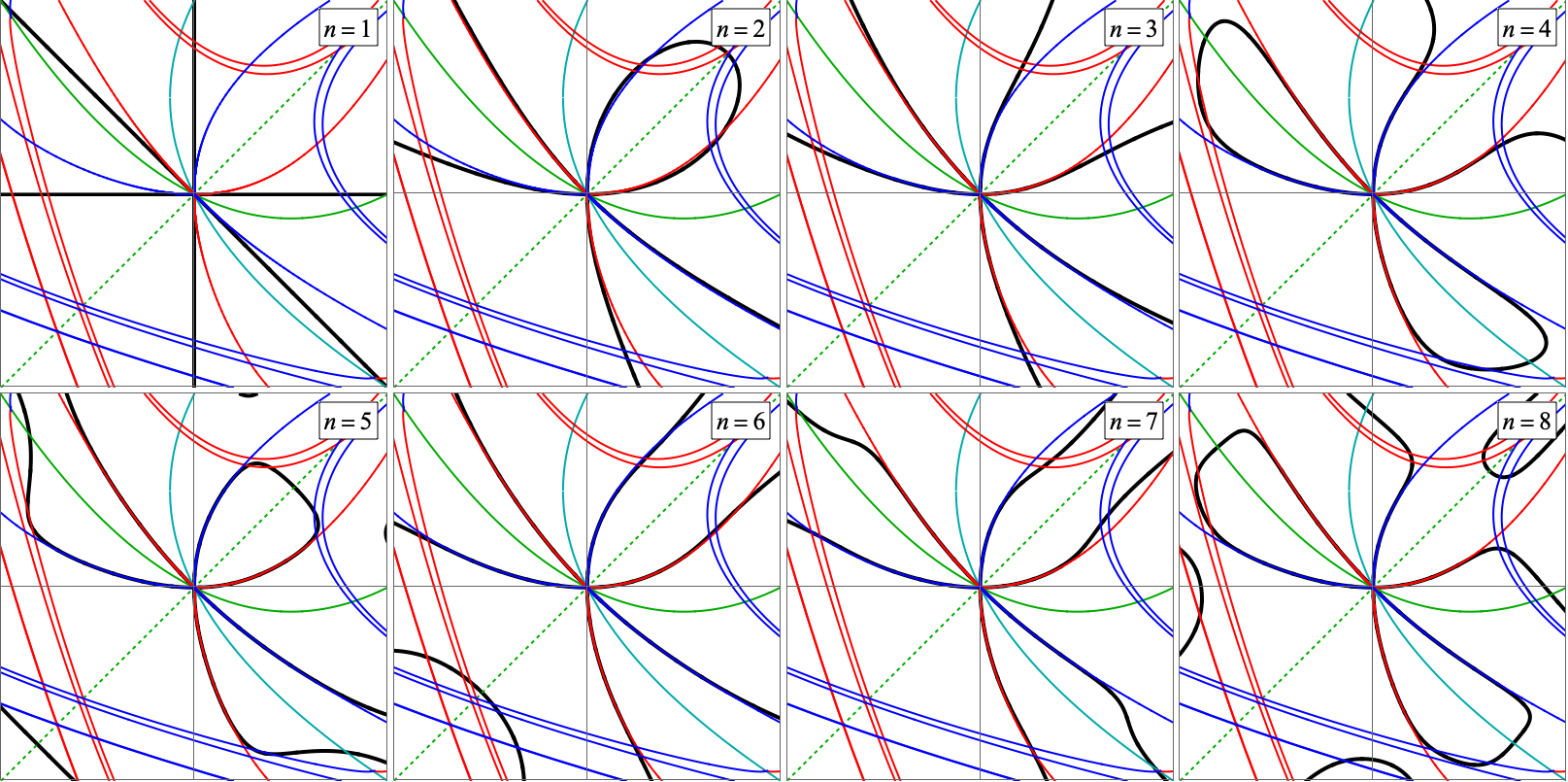}
    \caption{\label{fig:SXTr3}
    The top row illustrates some key invariant structures for
    the H\'enon quadratic map at the third-integer resonance
    $\nu_0 = 1/3$ ($a = 0$).
    Plot (c.1) highlights unstable fixed point
    $\z_\text{un}^{(1)}$, the fixed point at the origin that
    undergoes a touch-and-go bifurcation with an unstable
    3-cycle, and the neutral 3-cycle $\z^{(3)}$, which is on
    the verge of a period-doubling bifurcation.
    Linearized eigenvectors associated with stable (blue and purple)
    and unstable (red and pink) manifolds are shown, while the
    dashed black line indicates the direction of the eigenvector for
    points with neutral stability.
    Plot (c.2) provides a magnified view of one of the points in the
    parabolic 3-cycle $\z^{(3)}$, revealing some of the surviving
    KAM circles.
    Since the 3-cycle undergoes a period-doubling bifurcation, the horizontal-to-vertical aspect ratio is set to 1:10 for better visibility.
    Plot (c.3), shown on the same scale as (c.1), presents the stable
    and unstable manifolds associated with both fixed points, with a
    black rectangle marking the area for further comparison.
    The two bottom rows display the zero-level set of the approximate
    non-singular averaged invariant $\{\K^{(n)}_{a,b=1}\}[p,q] = 0$
    alongside the actual continuations of the stable and unstable
    manifolds from (c.3). Green curves represent the first (dashed)
    and second (solid) symmetry lines, while the cyan curve
    corresponds to the inverse of the second symmetry, $l_2^{-1}$.
    }\vspace{-0.25cm}
\end{figure*}

\newpage
\subsubsection{Third integer resonance, $\nu_0 = 1/3$}

Another important example is the system at the third-integer
resonance, $\nu_0 = 1/3$ ($a=-1$).
For a general force function with $b \neq 0$, even the first
twist coefficient $\tau_0$ becomes singular, resulting in
unstable motion near the origin.
The bottom row (c.) of Fig.~\ref{fig:RESs} illustrates the
phase space transformation for the simple quadratic H\'enon
map $f(p)=a\,p+p^2$.
When the system is slightly detuned from the resonance,
$\delta_r = \nu_r-\nu_0 \neq 0$, the invariant curves
of small-amplitude oscillations near the origin resemble
ellipses described by:
\begin{equation}
\label{math:CS13}    
    \cs[a=-1] = p^2 + p\,q + q^2.
\end{equation}

However, at the exact resonance, $\delta_r = 0$, the unstable
3-cycle, which otherwise defines the distinct triangular shape,
undergoes a touch-and-go bifurcation with the fixed point at the
origin.
Although the system exhibits local instability, some invariant
features persist, as highlighted in Fig.~\ref{fig:SXTr3}.
In plot (c.1), the system's fixed points and 3-cycles are
displayed along with the corresponding eigenvectors, obtained by
linearizing the Jacobian at each periodic orbit.
Plot (c.2) zooms into the region around the second 3-cycle
$\z^{(3)}$ (orange) at the point $(1,-1)$, showing some of the
surviving KAM circles.
Since $\z^{(3)}$ undergoes a period-doubling bifurcation, giving
rise to a 6-cycle when $\nu_0 > \nu_r$, it has neutral stability
with only one eigenvector, represented by the dashed black line.
Plot (c.3) further illustrates the extension of the eigenvectors
along the stable (blue and purple) and unstable (red and pink)
manifolds, originating from both the fixed point at the origin
and the unstable fixed point $\z^{(1)}_\mathrm{un}$.
As expected in chaotic systems, the manifolds exhibit homo- and
heteroclinic intersections, forming a complex web-like structure
often referred to as a {\it tangle}~\cite{sterling1999homoclinic}.

Since the perturbation expansion is constructed around the fixed
point at the origin, the associated stable and unstable manifolds
serve as exact invariant structures that the theory should ideally
capture.
While perturbation theory is not expected to accurately describe
regions dominated by tangles, it can still provide reliable
approximations in the vicinity of the origin where the structure
remains relatively simple (highlighted by the black square).
The subsequent two rows in Fig.~\ref{fig:SXTr3} present the
zero-level set of the approximate non-singular averaged
invariant:
\[
\{\K^{(n)}_{b=1}\}[p,q] = 0
\]
compared to the actual extensions of the stable and unstable
manifolds from plot (c.3).
While the stable and unstable manifolds attached to the fixed
point are each invariant in the sense of Eq.~(\ref{math:TG}),
their union forms a symmetric set that remains invariant under
Eqs.~(\ref{math:KK'}) and (\ref{math:RR'}).
As anticipated, within the chosen region, the zero-level set at
each successive order provides an increasingly accurate
approximation of the union of the manifolds.
The expansion effectively serves as a power series representation,
progressively refining the depiction of the invariant structure.

To gain a clearer understanding of the behavior near the origin
when $\delta_r = 0$, we examine the analytical expressions
obtained at the first few orders:
\[
\begin{array}{l}
\ds \{\K_{b=1}^{(1)}\} =-\Pi\,\Sigma\,\epsilon,   \\[0.25cm]
\ds \{\K_{b=1}^{(2)}\} =-\Pi\,\Sigma\,\epsilon - \left(
    \Pi^2 - \frac{5}{6}\frac{\cs^2}{2}
\right)\epsilon^2,                                  \\[0.25cm]
\ds \{\K_{b=1}^{(3)}\} =-\Pi\,\Sigma\,\epsilon - \left(
    \Pi^2 - \frac{\cs^2}{2}
\right)\epsilon^2\,-                                \\[0.25cm]
\ds\qquad -\, \left(
    \Pi + \frac{7}{9}\frac{\cs}{2}
\right)\Pi\,\Sigma\,\epsilon^3,
\end{array}
\]
\[
\begin{array}{l}
\ds \{\K_{b=1}^{(n \geq 4)}\} =-\Pi\,\Sigma\,\epsilon - \left(
    \Pi^2 - \frac{\cs^2}{2}
\right)\epsilon^2\,-                                \\[0.25cm]
\ds\qquad -\, \left(
    \Pi - \frac{\cs}{2}
\right)\Pi\,\Sigma\,\epsilon^3 + \ob(\epsilon^4).   
\end{array}
\]
The terms in $\K_{b=1}^{(n \geq 4)}$ converge up to $\epsilon^3$,
which accounts for the local instability near the origin.
The dominant first-order term, proportional to $\Pi\,\Sigma$,
represents the linearized contribution to the invariant.

Next, considering the case of a general smooth force function
given by $f(p) = a\,p + b\,p^2 + c\,p^3 + d\,p^4 + \ldots$.
The approximate invariant becomes
\[
\{\K_{a=-1}\} =
-b\,\Pi\,\Sigma\,\epsilon
-\left[
    b^2\,\Pi^2 - (b^2+c)\,\frac{\cs^2}{2}
\right]\epsilon^2 + \ob(\epsilon^3),
\]
as detailed in Appendix~\ref{secAPP:GSres}.
Notably, for small amplitudes, the invariant curves coincide with
ellipses described by Eq.~(\ref{math:CS13}) only if $b = 0$ and
$c \neq 0$;
in this scenario, the expansion begins with a $\cs^2$ term,
similar to the previously considered $\nu_0 = 1/4$ case.
However, when both $b,c = 0$ and $d \neq 0$, the system becomes
unstable once again, with the leading-order behavior described
by:
\[
\{\K_{a=-1;\,b,c=0}\} =
-d\,\Pi\,\Sigma\,\cs\,\epsilon^3 + \ob(\epsilon^4),
\]
consistent with the phenomenon of decapole instability in
accelerator physics~\cite{zolkinHenonSet}.

\subsection{\label{sec:Action}Action-angle variables}

Having constructed the approximate invariant and developed the
averaging procedure, we now turn our attention to how these results
can be connected to other perturbation theories, including Lie
algebra methods and related approaches.
In particular, we address the key question of how to derive the
approximate action-angle variables Eq.~(\ref{math:JTheta}) 
\[
\begin{array}{l}
    J' = J,                 \\[0.25cm]
    \theta' = \theta + 2\,\pi\,\nu(J),
\end{array}
\]
and how to obtain the twist coefficients or even the full
expression for the rotation number $\nu(J)$ when the series
converges:
\begin{equation}
\label{math:taus_1}
\nu(J) = \nu_0 + \tau_0 J +
\frac{1}{2!}\,\tau_1 J^2 +
\frac{1}{3!}\,\tau_2 J^3 + \ldots.
\end{equation}

This naturally leads us to the fundamental question: {\it ``Given
an invariant of motion and the mapping equations, how can one
determine the rotation number for a system that is exactly
integrable?''}
To our knowledge, this question was first articulated by Slava
Danilov during his Ph.D. studies at the Budker Institute of
Nuclear Physics, under the supervision of E.~Perevedentsev.
His insights were initially published and generalized in
\cite{zolkin2017rotation,nagaitsev2020betatron}, and later
extended to higher dimensions in~\cite{zolkin2024MCdynamicsII,
mitchell2021extracting}.

\begin{figure*}[th!]
    \includegraphics[width=\linewidth]{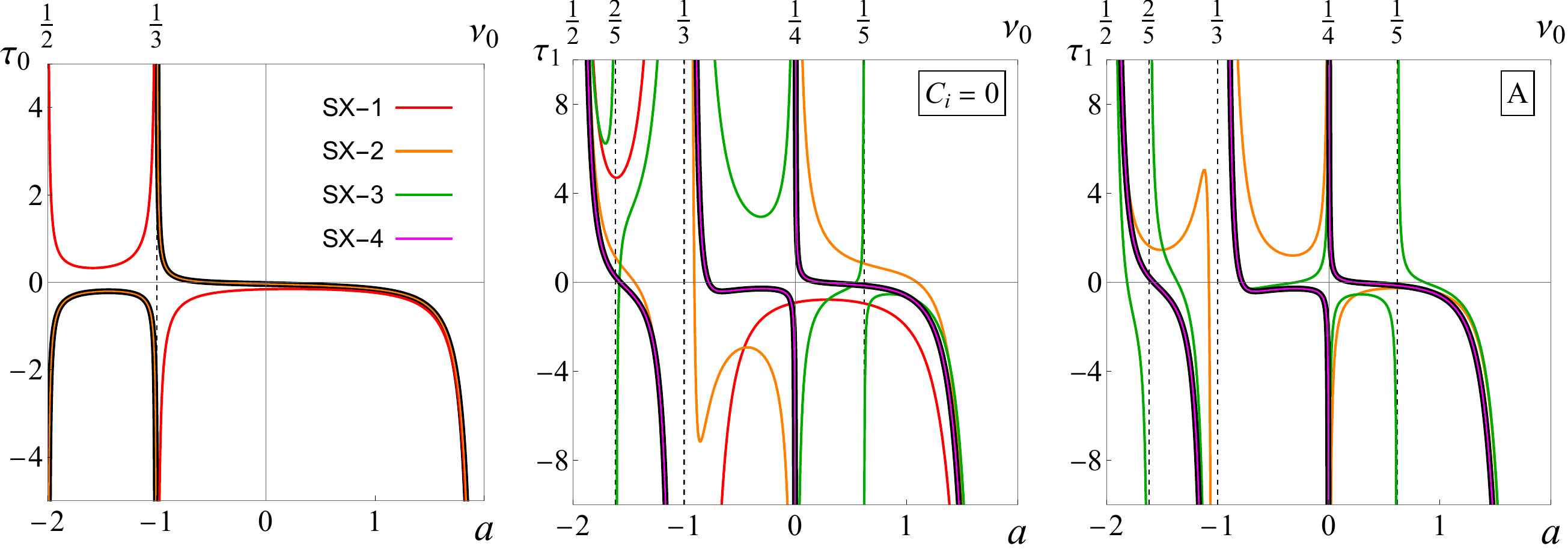}\vspace{-0.3cm}
    \caption{\label{fig:TwistConv}
    First two twist coefficients, $\tau_0(a)$ and $\tau_1(a)$,
    as functions of the trace for the H\'enon quadratic map
    $f(p) = a\,p + p^2$.
    The thick black curves represent results obtained using the
    Lie algebra method, Eqs.~(\ref{math:twist}) and
    (\ref{math:Twist12},\ref{math:Twist3}).
    Colored curves correspond to twist coefficients derived from
    the approximate invariants $\K^{(n)}_{b=1}[p,q]$ and are
    labeled SX-$n$ as indicated in the legend.
    Two sets of plots for $\tau_1$ are obtained: one using
    non-averaged invariants $\K^{(n)}_{b=1}[p,q]$ with all $C_i$
    set to zero, and another using the averaged invariants
    $\{\K^{(n)}_{b=1}\}[p,q]$, (A).
    Convergence is observed for $\tau_0$ at $n \geq 2$ and for
    $\tau_1$ at $n \geq 4$, irrespective of averaging.
    The complementary scale at the top indicates the bare rotation
    number, highlighting the relevant resonances.
    }
\end{figure*}

\noindent
\begin{theorem}[Slava Danilov]
Consider a symplectic map of the plane $\T:\z\mapsto\z'$, which
is integrable with an invariant (integral of motion) $\K[p,q]$
such that:
\[
\forall\,\z = (p,q)\in\mathbb{R}^2:\qquad
\K[\z'] - \K[\z] = 0.
\]
Theh, the Poincar\'e rotation number on a closed level set
$\K[p,q] = \const$, surrounding the fixed point, is given by the
ratio:
\[
\nu(\K) = \int_{\z_0}^{\z_0'}
    \left( \frac{\pd\K}{\pd p} \right)^{-1} \dd q
{\bigg /} \oint_{\K = \const}
    \left( \frac{\pd\K}{\pd p} \right)^{-1} \dd q.
\]
Alternatively, it can be expressed as:
\[
\nu(\K) = \frac{\dd J'(\K)}{\dd J(\K)}.
\]
Here, $J$ and $J'$ represent the action and partial action,
respectively, defined as:
\[
J(\K)   = \oint_{\K = \const} \frac{p\,\dd q}{2\,\pi},      \qquad\quad
J'(\K)  = \int_{\z_0}^{\z_0'} \frac{p\,\dd q}{2\,\pi}.
\]
In all cases, the integrals are taken along the invariant curve
$\K[p,q] = \const$.
$\int_{\z_0}^{\z_0'}$ refers to an integral over one iteration
of the map, remaining unaffected by the choice of initial point
$\z_0 =(p_0,q_0)$, whereas $\oint$ indicates a contour integral
encompassing the entire closed curve.
\end{theorem}

Therefore, at each order of the perturbation theory, within the
simply connected region around the fixed point, we can define an
approximate action and rotation number associated with the
invariant $\K^{(n)}[p,q]$:
\[
\begin{array}{l}
\ds J^{(n)}(\K^{(n)}) \equiv J(\K^{(n)}) = \oint_{\K^{(n)} = \const} \frac{p\,\dd q}{2\,\pi},\\[0.45cm]
\ds \nu^{(n)}(\K^{(n)}) = \frac{\dd J'(\K^{(n)})}{\dd J(\K^{(n)})}.
\end{array}
\]
By expanding both quantities in a power series of $\K^{(n)}$, we
obtain the desired expansion:
\[
\nu^{(n)}(J) =
    \nu_0 + \tau_0^{(n)} J +
    \frac{1}{2!}\,\tau_1^{(n)} J^2 +
    \frac{1}{3!}\,\tau_2^{(n)} J^3 +
    \ldots.
\]
Comparing this expansion with the coefficients $\tau_k$ derived
using the Lie algebra method,
we find that at any given order $n \geq 0$ and for any values of
$C_i$, the following general result holds:
\[
\tau_k^{(n)} = \tau_k
\quad
\text{for}
\quad
-1 \leq k \leq \lfloor n/2 \rfloor - 1,
\]
where $\tau_{-1} \equiv \nu_0$.
In other words, at each even order $n=2\,m$, all coefficients up
to $\tau_{m-1}^{(n)}$ converge to their corresponding exact values
$\tau_{m-1}$.

Fig.~\ref{fig:TwistConv} presents a comparison between the exact
coefficients $\tau_k$ (shown as black thick curves) and the
approximate coefficients $\tau_k^{(n)}$ (colored curves) obtained
from different orders of PT up to $n = 4$.
For illustration, we once again use the quadratic H\'enon map,
$f(p) = a\,p + p^2$.
The results for the approximate invariants $\K^{(n)}_{b=1}[p,q]$
are labeled as SX-$n$, representing their application to
horizontal dynamics in an accelerator with a thin sextupole (SX)
magnet~\cite{zolkin2024MCdynamics}.
The leftmost plot shows $\tau_0$, where the convergence of the
second-order approximation is evident, as expected;
here recall that SX-1 is independent of all $C_i$, while without
averaging, SX-2 and SX-3 are determined solely by $C_1$, and,
SX-4 depends on both $C_1$ and $C_2$.

\begin{figure*}[th!]
    \includegraphics[width=\linewidth]{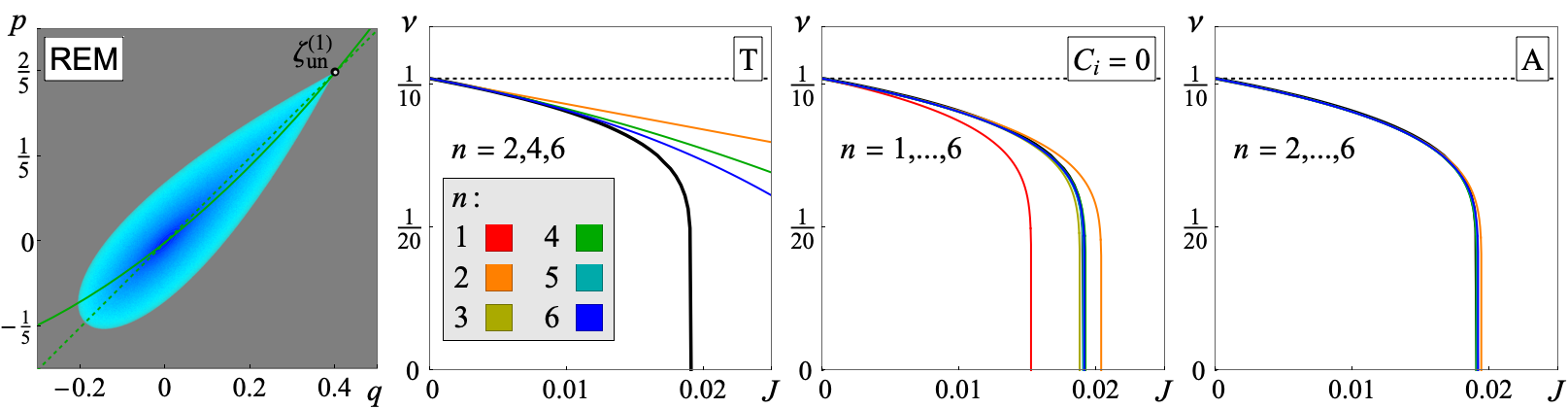}\vspace{-0.3cm}
    \caption{\label{fig:NuJ}
    The left plot displays the phase space of the H\'enon quadratic
    map $f(p) = a\,p + p^2$, at $a=8/5$, with colors representing
    the REM indicator.
    The three plots on the right compare approximations of $\nu(J)$
    (colored curves), obtained using different methods:
    the Taylor series (T) expansion of
    $\nu[J=0] = \nu_0 + \tau_0\,J + \ldots$ via the Lie algebra
    method (up to $J^3$, i.e., $n=6$),
    non-averaged invariants with $C_i = 0$,
    and the averaged invariants $\langle\K^{(n)}\rangle[p,q]$ (A).
    The black solid curve represents results obtained from
    numerical tracking, where $J$ is estimated as the area enclosed
    by an orbit over $2\,\pi$.
    }\vspace{-0.57cm}
\end{figure*}

\newpage
The fact that the {\it detuning} $\tau_0$, as defined in Eq.
(\ref{math:twist}), remains independent of $C_1$ has notable
consequences for maps with {\it typical} force functions, i.e.,
$b \neq 0$ or $c \neq 0$.
In particular, the second-order approximate invariant takes the
form
\[
\K^{(2)}[p,q] = \cs + \beta\,\Pi\,\Sigma\,\epsilon
    + \left(\alpha\,\Pi^2 + C_1\,\cs^2\right)\epsilon^2,
\]
where
\begin{equation}
\label{math:AlphaBeta}
\alpha = \frac{b^2}{r_3\,r_4} - \frac{c}{r_4},
\qquad\text{and}\qquad
\beta  =-\frac{b}{r_3}.
\end{equation}
Eq.~(\ref{math:AlphaBeta}) establishes a direct correspondence
between a typical map and an integrable McMillan map characterized
by the  symmetric invariant $\K^{(2)}[p,q]$ with $C_1 = 0$, and
the force
\[
    f_\mathrm{I}(p) =-\frac{\beta\,p^2-a\,p}{\alpha\,p^2 + \beta\,p + 1} =
    a\,p + b\,p + c\,p^2 + \ob(p^3).
\]
This correspondence guarantees that the force functions are
identical up to $\ob(\epsilon^3)$, which, in turn, ensures that
the rotation number expansions around the fixed point also match
up to the leading nonlinear term, $\nu = \nu_0 + \tau_0\,J$.
As shown in~\cite{zolkin2024MCdynamics,zolkin2024MCdynamicsIII},
integrable McMillan maps offer a robust framework for
understanding the behavior of typical mappings near the main
resonances, particularly $1/4$ and $1/3$, as well as integer
and half-integer resonances.
The analytical expressions and series expansions in these
references offer the necessary foundation to describe the lower
orders of PT.

The two final plots in Fig.~\ref{fig:TwistConv} illustrate the
behavior of the next twist coefficient $\tau_1$.
The middle plot shows the values of the second detuning with all
$C_i$ set to zero, while the last plot (labeled ``A'') presents
results obtained using the averaging procedure.
In both cases, the twist coefficient from the fourth-order
invariant (SX-4, magenta curve) aligns perfectly with the
prediction from the Lie algebra method (black curve).
While convergence is not expected for $n=2$, $3$ (orange and green
curves), careful inspection reveals that averaging improves
the results,

\newpage\noindent
 particularly near main resonances like the integer value
$\nu_0 = 0$ and $1/2,1/3,1/4$.

At this point, we have established that the converged terms
of the $\nu(J)$ series at $J=0$ yield the same results across
different methods --- whether derived from the Lie algebra
approach, perturbation theory (averaged or not), or direct
numerical experiments.
This naturally leads to the next important questions:
``Can $\nu^{(n)}(J)$ obtained from PT offer advantages
over the traditional expansion?'' and
``How does the choice of $C_i$ influence the results?''
While we explore the benefits of the averaging procedure in
the following sections, here we highlight one particularly
notable example.

Consider the quadratic H\'enon map slightly detuned from an
integer resonance, with $\nu_0 - \nu_r \approx 0.1$.
This case provides an example of {\it quasi-integrable} dynamics,
where the stable region is largely dictated by the invariant
manifolds associated with the second unstable fixed point,
$\z^{(1)}_\mathrm{un}$.
It not only has a relatively well-defined stability boundary but
also lacks visible island chains (though they do appear at smaller
scales), as seen in Fig~\ref{fig:NuJ}. 
The left plot shows the phase space using the REM
\cite{PANICHI201653, 10.1093/mnras/stx374, PhysRevE.107.064209}
indicator applied to a dense set of initial conditions.
Since our focus here is on enhancing the visualization compared
to previously used individual orbits, we omit the color scale
but note that more regular orbits appear in shades of blue to
cyan, while chaotic ones are marked in red and black (see also
Figures in the second part of this manuscript).

Besides a direct comparison of analytical approaches,
quasi-integrability allows for the extraction of both the action
(as the area enclosed by iterates of an orbit) and the rotation
number from numerical experiments
(black curve in all three plots).
For a detailed discussion of the numerical method used, refer to
Section 6.1.2 of~\cite{zolkin2024MCdynamics}.
A direct comparison of the curves reveals two key observations:
(i) PT provides a significantly more accurate approximation of
numerically computed $\nu(J)$ compared to the Taylor series (T),
and (ii) averaging (A) further improves accuracy.
This is evident, for example, when comparing the orange curves in
the last two plots.

\newpage
\section{\label{sec:Generalize}Possible generalizations}

\subsection{Symmetries}

A natural question for the curious reader is: what makes the first
symmetry special compared to the second, given that it is exactly
satisfied at every order?
This question becomes especially intriguing when one considers the
inverse mapping $\T^{-1}$, which reverses the symmetry ordering.

To explore this, we extend our analysis to a more general class of
maps~\cite{roberts1992revers}, related to the integrable asymmetric
McMillan map~\cite{IR2002II} and the QRT family
\cite{quispel1988integrable, quispel1989integrable}.
Consider a composite mapping constructed from two McMillan-form
transformations~(\ref{math:Ms}), each with a different force
function~\cite{roberts1992revers, IR2002II}:
\begin{equation}
\label{math:Mas}
\begin{array}{ll}
\T_a: \qquad    & q' =-q + f_1(p),           \\[0.25cm]
		      & p' =-p + f_2(q').
\end{array}
\end{equation}
This map is again reversible and symplectic for arbitrary force
functions $f_{1,2}$, and can be expressed as a composition of two
anti-area-preserving involutions, i.e.,
$\T_a = \R_2^\mathrm{(a)}\circ\R_1^\mathrm{(a)}$:
\[
\begin{array}{ll}
\R_1^\mathrm{(a)}:   & q' =-q + f_1(p),                   \\[0.25cm]
        & p' = p,
\end{array}
\qquad\quad
\begin{array}{ll}
\R_2^\mathrm{(a)}:   & q' = q,                   \\[0.25cm]
	& p' =-p + f_2(q).
\end{array}
\]
The corresponding symmetry lines are:
\[
l_1^\mathrm{(a)}:\quad q = f_1(p)/2
\qquad\text{and}\qquad
l_2^\mathrm{(a)}:\quad p = f_2(q)/2,
\]
where now both $l_1^\mathrm{(a)}$ and $l_2^\mathrm{(a)}$ are
defined by general functions.

We proceed with a perturbative expansion, starting again from a
general polynomial substitution.
Assuming a fixed point at the origin, $f_{1,2}(0)=0$, the
zeroth-order approximation yields the linearized invariant
(up to an overall factor $C_0$)
\[
\K_0^\mathrm{(a)}[p,q] = a_1\,p^2 - a_1\,a_2\,p\,q + a_2\,q^2,
\]
with the associated bare rotation number
\[
\nu_0 = \frac{1}{2\,\pi}\,\arccos\frac{\sigma}{2},
\qquad\qquad
\sigma = a_1\,a_2-2,
\]
where $\sigma$ is the trace of the Jacobian matrix evaluated at
the origin.

Before moving to higher orders, recall that for the symmetric
map, the approximate invariant $\K^{(n)}[p,q]$ had two distinct
contributions.
The first was a power series of the zeroth-order invariant
\[
C_0\,\cs + C_1\,\cs^2 + C_2\,\cs^3 + \ldots,
\]
defined up to an overall multiplier $C_0$ (seed), with
higher-order coefficients determined via the averaging procedure.
In contrast, the second set of terms contributes independently
and can be determined directly from the approximate invariance
condition.
These terms can be expressed in a matrix form:
\[
\begin{bmatrix}
1 \\[0.1cm] q \\[0.2cm] q^2 \\[0.2cm] q^3 \\[0.2cm] q^4 \\[0.2cm]
q^5 \\[0.1cm] \vdots
\end{bmatrix}
\cdot
\begin{bmatrix}
0       & 0       & \odot   & \oplus  & \odot   & \oplus & \odot  \\[0.25cm]
0       & \odot   & C_{2,1} & \odot   & C_{4,1} & \odot  &        \\[0.2cm]
\odot   & C_{1,2} & C_{2,2} & C_{3,2} & C_{4,2} &        &        \\[0.2cm]
\oplus  & \odot   & C_{2,3} & C_{3,3} &         &        &        \\[0.2cm]
\odot   & C_{1,4} & C_{2,4} &         &         &        &        \\[0.2cm]
\oplus  & \odot   &         &         &         &        &        \\[0.2cm]
\odot   &         &         &         &         &        &
\end{bmatrix}
\begin{bmatrix}
1   \\[0.1cm] p   \\[0.2cm] p^2 \\[0.2cm] p^3 \\[0.2cm]
p^4 \\[0.2cm] p^5 \\[0.1cm] \vdots
\end{bmatrix}.
\]
Here, per Proposition~\ref{prop1}, the coefficients satisfy the
symmetry $C_{i,j} = C_{j,i}$.
Additional Propositions~\ref{prop2} -- \ref{prop3}, dictate that
terms proportional to $p^{2k+1}$ and $q^{2k+1}$ vanish (marked
$\oplus$), while specific even-order terms $p^{2k}$, $p^{2k-1}q$,
$p\,q^{2k-1}$, $q^{2k}$ (marked $\odot$) are constrained and can
be set to zero to match the integrable symmetric McMillan mapping
for $C_{i>0} = 0$.

In the asymmetric case, the symmetry $C_{i,j} = C_{j,i}$ is lost.
Nonetheless, the structure of the approximate invariant still
follows the form
\[
C_0\,\K_0 + C_1\,\K_0^2 + C_2\,\K_0^3 + \ldots
\]
plus additional matrix, which now has less symmetry
\[
\begin{bmatrix}
1 \\[0.1cm] q \\[0.2cm] q^2 \\[0.2cm] q^3 \\[0.2cm] q^4 \\[0.2cm]
q^5 \\[0.1cm] \vdots
\end{bmatrix}
\cdot
\begin{bmatrix}
0       & 0       & \odot   & \oplus  & \odot   & \oplus & \odot\\[0.2cm]
0       & \odot   & C_{2,1} & \odot   & C_{4,1} & \odot  &      \\[0.2cm]
\odot   & C_{1,2} & C_{2,2} & C_{3,2} & C_{4,2} &        &      \\[0.2cm]
\oplus  & C_{1,3} & C_{2,3} & C_{3,3} &         &        &      \\[0.2cm]
C_{0,4} & C_{1,4} & C_{2,4} &         &         &        &      \\[0.2cm]
\oplus  & C_{1,5} &         &         &         &        &      \\[0.2cm]
C_{0,6} &         &         &         &         &        &
\end{bmatrix}
\begin{bmatrix}
1   \\[0.1cm] p   \\[0.2cm] p^2 \\[0.2cm] p^3 \\[0.2cm]
p^4 \\[0.2cm] p^5 \\[0.1cm] \vdots
\end{bmatrix}.
\]
Odd terms $p^{2k+1}$ and $q^{2k+1}$ again vanish.
For the even-order terms, only one of the two pairings can be
simultaneously eliminated: either the pair $p^{2k}$ and $p^{2k-1}q$
(marked with $\odot$), or the pair $p\,q^{2k-1}$ and $q^{2k}$.
Choosing the former set aligns the mapping with an asymmetric
McMillan map, characterized by the following force functions:
\[
f_1(p) =-\frac{\delta\,p^2 + \varepsilon\,p}{\alpha\,p^2+\beta\,p+\gamma},
\qquad
f_2(q) =-\frac{\beta\,q^2 + \varepsilon\,q}{\alpha\,q^2+\delta\,q+\kappa},
\]
with biquadratic invariant
\[
\K[p,q] = \alpha\,p^2q^2 + \beta\,p\,q^2 + \delta\,p^2q  + \gamma\,q^2 +
\varepsilon\,p\,q + \kappa\,p^2,
\]
and matched twist coefficient $\tau_0$ when $C_1 = 0$.
The coefficients (and the invariant itself) are defined up to a
common factor, let say $\varepsilon$, with relations:
\[
\begin{array}{ll}
\ds\beta  = \frac{\varepsilon}{a_1\,a_2}
    \frac{a_2^2\,b_1-a_1\,b_2}{a_1\,a_2-1},&
\ds\qquad\gamma =-\frac{\varepsilon}{a_1},\\[0.35cm]
\ds\delta = \frac{\varepsilon}{a_1\,a_2}
    \frac{a_1^2\,b_2-a_2\,b_1}{a_1\,a_2-1},&
\ds\qquad\kappa =-\frac{\varepsilon}{a_2},
\end{array}
\]
and
\[
\begin{array}{ll}
\alpha &\ds\!\!=\frac{\varepsilon}{a_1^2\,a_2}\left(
    a_2\,c_1-\frac{a_2^2\,b_1-a_1\,b_2}{a_1\,a_2-1}\,b_1
\right)     \\[0.35cm]
       &\ds\!\!=\frac{\varepsilon}{a_1\,a_2^2}\left(
    a_1\,c_2-\frac{a_1^2\,b_2-a_2\,b_1}{a_1\,a_2-1}\,b_2
\right).
\end{array}
\]

Proceeding to the evaluation of the two-indexed coefficients
$C_{i,j}$, the next-order contributions to the approximate
invariant take the form:
\[
\begin{array}{l}
\ds\K_1^\mathrm{(a)}[p,q] = -\frac{p\,q}{R_3}\left[
    (a_1^2\,b_2-a_2\,b_1)\,p + (a_2^2\,b_1-a_1\,b_2)\,q
\right],            \\[0.3cm]
\ds\K_2^\mathrm{(a)}[p,q] = \frac{p^2q^2}{R_3\,R_4}\left[
    L_{2,2} + L_{0,4}\left(1 - \frac{q}{a_1\,p}\right)^2
\right],            \\[0.3cm]
\cdots,
\end{array}
\]
where
\[
\begin{array}{l}
\ds L_{2,2} =
\frac{b_1 (a_2^2\,b_1 - a_1\,b_2) - a_2\,c_1\,R_3}{a_1}\,R_4,\\[0.35cm]
\ds L_{0,4} = a_1^3\,b_2^2 - a_2^3\,b_1^2 +
    (a_2^2\,c_1 - a_1^2\,c_2)\,R_3,
\end{array}
\]
and the resonant denominators that can be expressed in terms of
their symmetric counterparts as $R_k = r_k\left[a=\sigma\right]$.
For instance,
\[
R_3 = a_1\,a_2 - 1,\qquad\qquad
R_4 = a_1\,a_2 - 2.
\]
Analyzing these polynomials, we find that --- except in the
integrable asymmetric McMillan map case --- none of the symmetries
are exactly preserved at finite order.
Instead, $\K^{(n)}[p,q]$ respects both symmetries to an accuracy
matching the order, as given by Eq.~(\ref{math:Rn}).
The linear part of the symmetries is recovered in zeroth order,
the quadratic in first order, and so on.
This explains why the linear symmetry line $l_1$: $p=q$
(for the symmetric map $\mathrm{T}$) appears to be favored as one
proceeds to higher orders.

\subsection{Form of the map}

The same technique can be applied to a general symplectic map that
admits a polynomial representation:
\[
\begin{array}{cc}
\ds q' = A_{1,0}\,q + A_{0,1}\,p + A_{2,0}\,q^2 + A_{1,1}\,q\,p +
A_{0,2}\,p^2 + \ldots,\\[0.25cm] 
\ds p' = B_{1,0}\,q + B_{0,1}\,p + B_{2,0}\,q^2 + B_{1,1}\,q\,p +
B_{0,2}\,p^2 + \ldots.
\end{array}
\]
We briefly summarize the general procedure here.

\noindent$\bullet$
We begin by introducing a general polynomial ansatz for the
approximate invariant, where each $\K_m$ is a homogeneous
polynomial of degree $m+2$:
\[
\K^{(n)}[p,q] = \K_0 + \epsilon\,\K_1 + \ldots + \epsilon^n\,\K_n,
\]
which we require to be conserved up to order $\ob(\epsilon^{n+1})$
\[
\Rs_n = \K^{(n)}[p',q'] - \K^{(n)}[p,q] = 
    \overline{\Rs_n}\,\epsilon^{n+1} +
    \ob(\epsilon^{n+2}).
\]
The lowest-order terms of the approximate invariant --- also
obtainable through standard linearization --- are defined up
to a common multiplicative factor
\[
\K_0[p,q]=C_0\left[
    A_{0,1}\,p^2+(A_{1,0}-B_{0,1})\,p\,q-B_{1,0}\,q^2
\right].
\]
The seed coefficient $C_0$ can be set to unity, or to a specific
value chosen to eliminate resonant denominators.
Alternatively, choosing $C_0 = \beta^*/A_{0,1}$, where $\beta^*$
is the beta-function at a given location in an accelerator,
recovers the conventional Courant-Snyder invariant,
$C_0\,\K_0 = \mathrm{em}$, where $\mathrm{em}$ is the beam
emittance, defining the area occupied by the beam in phase space.

\noindent$\bullet$
Due to under-determinacy, the invariant (after all two-indexed
coefficients $C_{i,j}$ have been determined) is known only up to
a series
\[
    C_0\,\K_0 + C_1\,\K_0^2 + \ldots.
\]
Nevertheless, the approximate invariance condition holds to the
required order, and the twist coefficient $\tau_k$ converges at
order $n=2\,k+2$.

\noindent$\bullet$
To determine the coefficients $C_i$ (for $i>0$), we use an
averaging procedure.
Using the eigenbasis of the Jacobian $(q,p)\rightarrow(Q,P)$,
the linear part of the map is transformed into a pure rotation:
\[
\begin{array}{cc}
\ds Q' = Q\,\cos(2\,\pi\,\nu_0) - P\,\sin(2\,\pi\,\nu_0) + \ldots,\\[0.25cm] 
\ds P' = Q\,\sin(2\,\pi\,\nu_0) + P\,\cos(2\,\pi\,\nu_0) + \ldots,
\end{array}
\]
and the linearized invariant becomes:
\[
\K_0[P,Q] = P^2 + Q^2.
\]
In accelerator physics, this transformation is typically achieved
through the use of Floquet variables.
Switching to polar coordinates:
\[
    Q = \rho\cos\psi,
    \qquad
    P = \rho\sin\psi,
\]
the coefficients $C_k$ are determined by solving the system:
\[
    \frac{\dd}{\dd C_k}\,I_n = 0,
\]
which minimizes the integral
\[
I_n = \int_0^{2\pi} \overline{\Rs_n}^2[\rho,\psi] \,\dd\psi.
\]

\noindent$\bullet$
{\bf Note}: If the map is not explicitly constructed from a
sequence of symplectic transformations, one must ensure
symplecticity by enforcing conditions such as:
\[
\begin{array}{l}
\ds A_{1,0}\,B_{0,1} - B_{1,0}\,A_{0,1} = 1,    \\[0.25cm]
\ds A_{1,1}\,B_{0,1} + 2\,A_{1,0}\,B_{0,2} -
 2\,A_{0,2}\,B_{1,0} - A_{0,1}\, B_{1,1} = 0,   \\[0.25cm]
\ds A_{1,1}\,B_{1,0} + 2\,A_{0,1}\,B_{2,0} -
 2\,A_{2,0}\,B_{0,1} - A_{1,0}\,B_{1,1} = 0,    \\[0.25cm]
\cdots.
\end{array}
\]

\subsection{Dimensions}

Finally, we note that this approach can be extended to systems of
higher dimensionality.
For instance, in a 4D symplectic map involving two coordinates and
momenta $(x,p_x,y,p_y)$, and assuming linearly decoupled dynamics,
one begins with horizontal and vertical quadratic invariants:
``horizontal'' $\K_0^{(x)}[x,p_x]$ and
``vertical'' $\K_0^{(y)}[y,p_y]$.
Nonlinear coupling is then recovered in higher-order terms such as
$\K_1^{(x,y)}[x,p_x,y,p_y]$.
In this case, trajectories are no longer confined to a single level
set, but instead lie on the intersection of two 4D hypersurfaces.
This complicates the analysis but still allows for accurate estimates.
We leave a detailed discussion to a separate article.

\section{\label{sec:Conclusion} Conclusions and discussion}

In closing, we would like to comment on the remarkable robustness
of particle accelerators to generic perturbations in their
nonlinear optics, and the resulting stability this imparts.

Accelerator lattices are often considered ``poorly'' optimized
when it comes to nonlinear elements.
This characterization stems from multiple sources.
Operationally, not all nonlinear magnets are always available due
to technical issues or failures.
As accelerator complexes grow older and more expansive, such
limitations tend to accumulate.
Compounding this, there are numerous sources of parasitic nonlinear
fields, arising from fringe effects, magnet imperfections, or design
flaws.
At best, these imperfections are carefully measured after
construction and partially corrected.
Nevertheless, the cumulative impact is rarely negligible.
And yet, these machines continue to deliver beams reliably,
extracting maximum performance from limited resources.

The term {\it stability} deserves clarification.
On one hand, particles in accelerators do exhibit instabilities ---
often due to collective effects.
As beam intensity increases, so too does the influence of interactions
among particles and with the environment.
Energy growth, likewise, leads to intra-beam scattering.

On the other hand, at lower beam intensities and energies ---
where collective effects are suppressed and nonlinear magnetic
fields dominate --- the dynamics of single particles become the
primary concern.
In this regime, the region of stability is typically defined either
by a singular resonance (with clear stable and unstable manifolds,
as seen in Fig.~\ref{fig:NuJ}), or by the overlap of multiple
resonances, which results in island chains and chaotic trajectories.

This distinction becomes evident when examining maps in an
extended parameter space.
In the second part of this article, we use isochronous stability
diagrams, where singular resonances yield well-defined boundaries,
in contrast to the ``ripped'' fractal edges created by overlapping
resonances
(see also \cite{zolkinHenonSet} for a detailed discussion).

A central question then arises: how can one estimate the region of
stability?
For singular resonances, the stability region is, roughly speaking,
bounded by the distance to the associated unstable fixed point or
$n$-cycle.
However, far from any singular resonance
(in the space of parameters), estimating the stability region
becomes highly nontrivial.
What is particularly striking is that resonance overlap tends to
occur only at relatively large amplitudes.
This allows for a surprisingly large domain of quasi-integrable
motion to persist --- and, in some cases, even quasi-linear
behavior --- where chaos indicators such as SALI/GALI or REM vary
very little.
Similar effects have been observed in piecewise linear systems
\cite{ZKN2023PolI,ZKN2024PolII}.

We suggest that this persistence of stability (complementing
general results from KAM theory) can, at least in part, be
understood through the presence of a low-order, nonlinear,
integrable map hidden near the origin.
This is akin to how symmetric and asymmetric mappings approximate transformations of the corresponding form.
Even for a general map, the existence of two symmetry lines
intersecting at the fixed point (even if not known explicitly)
supports the construction of approximate invariants.

In higher-dimensional systems, such as particle accelerators
governed by 6D symplectic maps derived from time-dependent
Hamiltonians, the planar approximation can provide remarkably
accurate and practical results.
In practice, many particle beams are {\it flat}, meaning their
horizontal size is much larger than the vertical,
$\sigma_x \gg \sigma_y$.
In such settings, horizontal invariants provide highly accurate
estimates of dynamic aperture limits imposed by singular resonances.
Another notable example is resonant extraction, which exploits a
singular resonance to channel streams of particles out of the
machine.
In this context, approximate invariants once again provide fast
and effective estimates --- similar to the classical isolated
resonance approach.

The third part of this manuscript contains a variety of
examples based on general symplectic mappings derived from
realistic accelerator lattices across the FermiLab complex,
demonstrating the method's practical efficiency in
real-world applications.

\section{Acknowledgments}

The authors would like to thank Taylor Nchako (Northwestern
University) for carefully reading this manuscript and for her
helpful comments.
S.N. work is supported by the U.S. Department of Energy,
Office of Science,
Office of Nuclear Physics under contract DE-AC05-06OR23177.
I.M. acknowledges that his work was partially supported by the
Ministry of Science and Higher Education of the Russian Federation
(project FWUR-2025-0004).
S.K. is grateful to his supervisor, Prof. Young-Kee Kim
(University of Chicago), for her valuable mentorship and continuous
support.

\appendix

\newpage
\onecolumngrid
\section{\label{secAPP:GS}Map in McMillan form with analytic
force function}

In this appendix, we summarize some analytical results for the
map in McMillan form~(\ref{math:Ms}), assuming a force function
that admits a Taylor series expansion:
\begin{equation}
\label{math:fGeneral}    
f(\epsilon\,p)/\epsilon = a\,p
    + \epsilon\, b\,p^2 + \epsilon^2 c\,p^3
    + \epsilon^3 d\,p^4 + \epsilon^4 e\,p^5
    + \epsilon^5\,\mathrm{f}\,p^6 + \epsilon^6 g\,p^7 + \ldots,
\end{equation}
where $f(0) = 0$, ensuring a fixed point at the origin.

\vspace{-0.2cm}
\subsection{\label{secAPP:GStwist}Twist expansion}

Expanding the rotation number in terms of the action variable up
to the third order in $J^3$ (corresponding to $\epsilon^6$)
\[
\nu[J=0] = \nu_0 + \tau_0 J +
\frac{1}{2!}\,\tau_1 J^2 +
\frac{1}{3!}\,\tau_2 J^3 + \ldots,
\]
for the force function in Eq.~(\ref{math:fGeneral}), the twist
coefficients are given by:
\begin{equation}
\label{math:Twist12}
\begin{array}{l}
\ds \frac{2\,\pi\,\nu_0}{0!} = \arccos[a/2],
\qquad\qquad\qquad\qquad\qquad\qquad\qquad\qquad\qquad\qquad
\frac{2\,\pi\,\tau_0}{1!} =
\frac{3}{r_1\,r_2}\left[
    c-\frac{2}{3}\,\frac{1+2\,a}{r_1\,r_3}\,b^2
\right],                                    \\[0.45cm]
\ds \frac{2\,\pi\,\tau_1}{2!} =
\frac{3}{(-r_1\,r_2)^{5/2}\,r_4}\left[
    \left(1-\frac{9}{2}\,a^2\right)c^2 +
    4\,\frac{t_{11}}{r_1\,r_3^2}\,b^2 c -
    \frac{2}{3}\,\frac{t_{12}}{r_1^2\,r_3^3}\,b^4
\right] +
\frac{2}{(-r_1\,r_2)^{3/2}}\left[
    6\,\frac{1+3\,a}{r_1\,r_3}\,b\,d -5\,e
\right],
\end{array}
\end{equation}
and for $\mathrm{f} = g = 0$
\begin{equation}
\label{math:Twist3}
\begin{array}{l}
\ds \!\!\!\!\!\!\!\!\!\!\!\!\!\!\!
\frac{2\,\pi\,\tau_2}{3!} =
-\frac{10}{(r_1\,r_2)^4\,r_4^2}\left[
    6\left(1+\frac{3}{2}\,a^4\right)c^3 -
    2\,\frac{t_{21}}{r_1\,r_3^3\,r_5}\,b^2 c^2 +
    \frac{t_{22}}{r_1^2\,r_3^4\,r_5}\,b^4 c -
    \frac{2}{3}\,\frac{t_{23}}{r_1^3\,r_3^5\,r_5}\,b^6
\right]            \\[0.45cm]
\ds \,+\,
\frac{4}{(r_1\,r_2)^3\,r_3^2\,r_4}\left[
    \left\{
        r_2\,r_3\,r_4\,t_{24}\,d -
        10\,\frac{t_{25}}{r1}\,b\,c +
        10\,\frac{t_{26}}{r_1^2\,r_3}\,b^3
    \right\}\frac{d}{r_5} +
    5\left\{ 2\,r_3^2\,(3\,a^2-1)\,c-\frac{t_{27}}{r_1}\,b^2 \right\}e
\right],
\end{array}
\end{equation}
where
\[
\begin{array}{l}
t_{11} = 6\,a^4 + 12\,a^3 + 6\,a^2 + 4\,a + 1,              \\[0.2cm]
t_{12} = 16\,a^5 + 32\,a^4 + 27\,a^3 + 37\,a^2 + 14\,a - 6, \\[0.2cm]
t_{21} = 54\,a^9 + 234\,a^8 + 339\,a^7 + 222\,a^6 + 164\,a^5
        + 54\,a^4 - 152\,a^3 - 94\,a^2 - 6\,a + 10,         \\[0.2cm]
t_{22} = 136\,a^{10} + 600\,a^9 + 984\,a^8 + 1144\,a^7
        + 1411\,a^6 + 455\,a^5 - 1007\,a^4 - 496\,a^3
        + 212\,a^2 + 88\,a - 8,                             \\[0.2cm]
t_{23} = 56\,a^{11} + 240\,a^{10} + 440\,a^9 + 764\,a^8
        + 1110\,a^7 + 360\,a^6 - 632\,a^5 - 129\,a^4
        + 292\,a^3 - 44\,a^2 - 120\,a - 24,                 \\[0.2cm]
t_{24} = 18\,a^3 + 20\,a^2 - 16\,a - 3,                     \\[0.2cm]
t_{25} = 18\,a^6 + 48\,a^5 + 20\,a^4 - 14\,a^3 + 3\,a^2
        - 4\,a - 2,                                         \\[0.2cm]
t_{26} = 20\,a^7 + 56\,a^6 + 38\,a^5 + 22\,a^4 + 30\,a^3
        - 25\,a^2 - 22\,a + 4,                              \\[0.2cm]
t_{27} = 20\,a^4 + 40\,a^3 + 15\,a^2 + 10\,a + 4.
\end{array}
\]

\vspace{-0.2cm}
\subsection{\label{secAPP:GSapprox}Approximate invariant of motion}

The approximate non-averaged invariant up to the fourth order in
$\epsilon$ is given by:
\[
\begin{array}{l}
\ds \K^{(4)}[p,q] =
    \cs -
    \frac{b}{r_3}\,\Pi\,\Sigma\,\epsilon +
    \frac{b^2-r_3\,c}{r_3\,r_4}\,\Pi^2\epsilon^2 +
    C_1\,\cs^2\epsilon^2 -
    \frac{\mathcal{T}_0}{r_3\,r_4\,r_5}\,
    \left[
        \Pi - \frac{\cs}{r_3}
    \right]\Pi\,\Sigma\,\epsilon^3 -
    C_1\frac{2\,b}{r_3}\,\Pi\,\Sigma\,\cs\,\epsilon^3   \\[0.55cm]
\ds \qquad +\,
\frac{1}{r_3^2\,r_4\,r_5}\,\left[
    \frac{
        (2+r_2\,r_4)\,b^4 -
        \mathcal{T}_1\,b^2\,c +
        r_3^2\,\left\{r_4\,(r_3+r_4)\,b\,d + r_5\,(c^2-r_4\,e)\right\}
    }{r_3\,r_6}
    \left( \Pi - \frac{\cs}{r_4} \right)
    -
    2\,b\,\mathcal{T}_0\,\frac{\cs}{r_4}
\right]\Pi^2\epsilon^4                                  \\[0.55cm]
\ds \qquad +\,
    \frac{C_1}{r_3^2} \left[
        r_2\,b^2\,\Pi + \frac{(2+3\,a)\,b^2 - 2\,r_3^2\,c}{r_4}\,\cs
    \right] \Pi^2\epsilon^4 +
    C_2\,\cs^3\epsilon^4,
\end{array}
\]
where
\[
\mathcal{T}_0 = b^3-(r_3+r_4)\,b\,c+r_3\,r_4\,d
\qquad\qquad\text{and}\qquad\qquad
\mathcal{T}_1 = 3\,a^3 + 7\,a^2 + 6\,a +1.
\]

\subsection{\label{secAPP:GSav}Averaging procedure}

Performing the averaging procedure for orders $n=2$ and $n=3$,
where we have a single coefficient $C_1$, provides:
\[
\begin{array}{l}
\ds C_1^{(2)} =
\frac{5}{4}\,\frac{\mathcal{T}_0}{r_1\,r_2\,r_3\,r_4\,b},   \\[0.45cm]
\ds C_1^{(3)} =
\frac{
    T_1\,b^6 - T_2\,b^4\,c + r_3\,r_4\,T_3\,b^3\,d +
    r_3\,(T_4\,c^2-7\,T_5\,r_3\,r_4\,r_5\,e)\,b^2 -
    2\,r_1\,r_3^2\,r_4\,T_6\,b\,c\,d +
    14\,r_1\,r_3^3\,r_5\,(r_4\,e - c^2)\,c
}{
r_2\,r_3\,r_4\,r_5\left(
    T_7\,b^4 - 4\,r_1\,r_3\,T_8\,b^2\,c + 10\,r_1^2\,r_3^2\,c^2
\right)
}.
\end{array}
\]
At higher orders, the equations become more complex, and we
encounter multiple coefficients.
For clarity, we provide the results for two examples that are
extensively used in our article: the generalized cubic map,
$f(p) = a\,p + b\,p^2 + c\,p^3$:
\[
\begin{array}{l}
\ds C_1^{(4)} =-\frac{1}{r_3\,r_4\,r_5\,r_6}\,
\frac{
    r_4\,U_1\,b^6 - U_2\,b^4\,c +
    r_3\,U_3\,b^2\,c^2 - 2\,r_3^2\,U_4\,c^3}
{U_{10}\,b^4 - 4\,r_3\,U_{11}\,b^2\,c + 4\,r_3^2\,U_{12}\,c^2}, \\[0.45cm]
\ds C_2^{(4)} =-\frac{2}{3}\,\frac{1}{r_1\,r_2\,r_3\,r_4\,r_5\,r_6}\,
\frac{
    2\,r_4\,(3\,r_4+2)\,U_5\,b^8 - U_6\,b^6\,c + U_7\,b^4\,c^2 -
    2\,r_3\,U_8\,b^2\,c^3 + 4\,r_3^2\,U_9\,c^4}
{U_{10}\,b^4 - 4\,r_3\,U_{11}\,b^2\,c + 4\,r_3^2\,U_{12}\,c^2},
\end{array}
\qquad [d,e,\mathrm{f},g=0]
\]
and the odd-force mapping
$f_\mathrm{odd}(p) =a\,p+c\,p^3+e\,p^5+g\,p^7+i\,p^9+\ldots$,
which includes, in particular, the simple cubic map
$f(q) = a\,p + c\,p^3$
(remember, that for this case, the averaged coefficients must be
derived from the $\overline{\overline{\Rs_n}}$, rather than from
$\overline{\Rs_n}$ as used in the equations above):
\[
\begin{array}{l}
\ds C_1^{(4\mathrm{-odd})} = \frac{1}{r_4\,(r_6\,r_3)}\,
\frac{
    5\,V_1\,c^5 - 6\,r_4\,V_2\,c^3\,e + 3\,r_4^2\,(r_6\,r_3)\,V_3\,c^2\,g -
    8\,(r_2\,r_1)\,r_4^2\,V_4\,c\,e^2 + 144\,(r_2\,r_1)\,r_4^3\,(r_6\,r_3)\,e\,g
}{V_7\,c^4 + 64\,r_2\,r_1\,r_4\,V_8\,c^2\,e + 56\,(r_2\,r_1)^2\,r_4^2\,e^2},
\qquad\,\!\!\!\!\!\!\\[0.45cm]
\ds C_2^{(4\mathrm{-odd})} =\frac{-28}{(r_2\,r_1)\,r_4\,(r_6\,r_3)\,c}\,
\frac{
    \left[V_5\,c^2 + 4\,(r_2\,r_1)\,r_4\,e\right]
    \left[V_6\,c + r_4\,(r_6\,r_3)(2\,c^2-r_4\,e)\,g\right]
}{V_7\,c^4 + 64\,r_2\,r_1\,r_4\,V_8\,c^2\,e + 56\,(r_2\,r_1)^2\,r_4^2\,e^2},
\end{array}
\]
where
\[
\begin{array}{l}
\,\!\!\!\begin{array}{ll}
T_1 = 13\,a^3-47\,a^2-80\,a+34,         &\qquad\qquad
V_1 = 25\,a^4+103\,a^2-50,              \\[0.2cm]
T_2 = 57\,a^4-98\,a^3-365\,a^2-63\,a+93,&\qquad\qquad
V_2 = 3\,a^4+282\,a^2-160,              \\[0.2cm]
T_3 = 20\,a^3-75\,a^2-122\,a+69,        &\qquad\qquad
V_3 = 31\,a^2+130,                      \\[0.2cm]
T_4 = 69\,a^4-37\,a^3-282\,a^2-41\,a+87,&\qquad\qquad
V_4 = 43\,a^2-28,                       \\[0.2cm]
T_5 = a-5,                              &\qquad\qquad
V_5 = 3\,a^2+10,                        \\[0.2cm]
T_6 = 19\,a^2+21\,a-13,                 &\qquad\qquad
V_6 = 2\,a^3\,e^2-6\,a^2\,c^2\,e+3\,a\,c^4+c^2\,e,\\[0.2cm]
T_7 = 5\,a^2-34\,a+69,                  &\qquad\qquad
V_7 = 29\,a^4+140\,a^2+500,             \\[0.2cm]
T_8 = 3\,a-13,                          &\qquad\qquad
V_8 = a^2+5,
\end{array}                                         \\[2.3cm]
\text{and}                                          \\[0.2cm]
U_1 = (a^3+2\,a^2+2\,a+2)(15\,a^3+30\,a^2+8\,a-18), \\[0.2cm]
U_2 = 72\,a^8+299\,a^7+549\,a^6+634\,a^5+380\,a^4-
    131\,a^3-288\,a^2-60\,a+18,                     \\[0.2cm]
U_3 = 82\,a^8+324\,a^7+655\,a^6+764\,a^5+168\,a^4-
    507\,a^3-344\,a^2+8\,a+36,                      \\[0.2cm]
U_4 = 12\,a^8+58\,a^7+143\,a^6+122\,a^5-78\,a^4-
    162\,a^3-56\,a^2+16\,a+9,                       \\[0.2cm]
U_5 = 15\,a^3+22\,a^2+12\,a+20,                     \\[0.2cm]
U_6 = 468\,a^6+1278\,a^5+1518\,a^4+1557\,a^3+
    962\,a^2+108\,a-20,                             \\[0.2cm]
U_7 = 809\,a^7+2838\,a^6+4620\,a^5+5131\,a^4+
    3423\,a^3+652\,a^2-356\,a-80,                   \\[0.2cm]
U_8 = 291\,a^7+987\,a^6+1697\,a^5+1738\,a^4+
    616\,a^3-407\,a^2-342\,a-50,                    \\[0.2cm]
U_9 = (4\,a^3+11\,a^2+20\,a+10)
    (9\,a^4+8\,a^3-5\,a^2-6\,a-1),                  \\[0.2cm]
U_{10} = 11\,a^6+32\,a^5+40\,a^4+60\,a^3+80\,a^2+
    64\,a+36,                                       \\[0.2cm]
U_{11} = 3\,a^6+12\,a^5+26\,a^4+41\,a^3+56\,a^2+
    52\,a+18,                                       \\[0.2cm]
U_{12} = a^6+4\,a^5+10\,a^4+26\,a^3+48\,a^2+36\,a+9.
\end{array}
\]

Additionally, upon request, we can provide a \texttt{Wolfram Mathematica}
file with simplified expressions for the general force mapping
up to $\epsilon^6$, as well as for the mixed force (generalized
cubic) $f_\mathrm{MX}(p) = a\,p + b\,p^2 + c\,p^3$ up to
$\epsilon^8$, the quadratic map up to $\epsilon^{10}$, and the
simple cubic map ($b=0$) up to $\epsilon^{16}$.

\subsection{\label{secAPP:GSres}
Normal form of approximate invariant for low order resonances}

Consider a general map in McMillan form with a force function
\[
f(q) = a\,q + b\,q^2 + c\,q^3 + d\,q^4 + e\,q^5 + f\,q^6 + \ldots.
\]
Below, we present the converged lower-order terms of the averaged
invariant $\langle \K \rangle[p,q]$ at selected primary resonances.
For the third-integer resonance $\nu_0 = 1/3$ (i.e., $a=-1$), we
obtain:
\[
\begin{array}{ll}
b \neq 0: &\ds
-b\,\Pi\,\Sigma\,\epsilon
-\left[
    b^2\,\Pi^2 - (b^2+c)\,\frac{\cs^2}{2}
\right]\epsilon^2
-b\,\left[
    (b^2+c)\,\Pi - \left(
        b^2-c - 2\,\frac{d}{b}
    \right)\frac{\cs}{2}
\right]\Pi\,\Sigma\,\epsilon^3 + \ob(\epsilon^4),       \\[0.45cm]
b = 0: &\ds
    c\,\frac{\cs^2}{2}\,\epsilon^2 -
    d\,\Pi\,\Sigma\,\cs\,\epsilon^3 +
    \frac{1}{2}\left[
        (c^2+e)\,\Pi^3 +
        (3\,c^2+e)\,\Pi^2\,\cs -
        \frac{3\,c^2-2\,e}{3}\,\cs^3
    \right]\epsilon^4                                   \\[0.45cm]
&\ds\qquad - \left[
        c\,d\,\Pi^2 -
        (c\,d-2\,f)\,\frac{\cs^2}{2}
    \right]\Pi\,\Sigma\,\epsilon^5
    + \ob(\epsilon^6).
\end{array}
\]
For the fourth-integer resonance $\nu_0 = 1/4$ (i.e., $a=0$), we
find:
\[
\begin{array}{ll}
b \neq 0: &\ds
\left[
    (b^2-c)\,\Pi^2 + c\,\frac{\cs^2}{2}
\right]\epsilon^2 +
b\,\left[
    (b^2-c)\,\Pi - b^2\,\cs
\right]\Pi\,\Sigma\,\epsilon^3                          \\[0.45cm]
&\ds +\,
\left[
    (2\,b^4-c^2)\,\Pi^3 -
    \left(b^4 - \frac{3}{2}\,b^2\,c - b\,d + e\right)\,\Pi^2\,\cs +
    (b^4+e)\,\frac{\cs^3}{3}
\right]\epsilon^4                                       \\[0.45cm]
&\ds +\,
b\,\left[
    \left(b^4 - c^2 - c\,\frac{d}{b} + e\right)\,\Pi^2 -
    \left(b^4 - b^2c + c^2 - 2\,b\,d + e\right)\,\Pi\,\cs -
    (b^2\,c - c^2 + 2\,b\,d)\,\cs^2
\right]\Pi\,\Sigma\,\epsilon^5                          \\[0.45cm]
&\ds +\,\ob(\epsilon^6),                                \\[0.45cm]
b = 0: &\ds
-c\,\left[
    \Pi^2 - \frac{\cs^2}{2}
\right]\epsilon^2 -
\left[
    c^2\,\Pi^3 + e\,\Pi^2\,\cs - e\,\frac{\cs^3}{3}
\right]\,\epsilon^4 -
c\,d\,\Pi^3\,\Sigma\,\epsilon^5 + \ob(\epsilon^6).
\end{array}
\]
For the resonance $\nu_0 = 1/6$ (i.e., $a=1$), the result is:
\[
\begin{array}{ll}
b \neq 0: &\ds
(b^2+c)\,\frac{\cs^2}{2}\,\epsilon^2 -
b\,(b^2+c)\,\frac{\Pi\,\Sigma\,\cs}{2}\,\epsilon^3 +
\ob(\epsilon^4),                                        \\[0.45cm]
b,c = 0: &\ds
-\frac{e}{2}\left[
    \Pi^3- \left(\Pi^2 + \frac{2}{3}\,\cs^2\right)\,\cs
\right]\epsilon^4
+ \ob(\epsilon^6).
\end{array}
\]
For the fifth-order resonances $\nu_0 = 1/5$ and $\nu_0 = 2/5$
we obtain the corresponding averaged invariant
\[
\begin{array}{ll}
d,e = 0: &\ds
    a_r\,\frac{2\,b^2 + 3\,c}{2}\,\cs^2\,\epsilon^2 -
    b\left[
    \left\{b^2 - (2\,a_r + 1)\,c\right\}\Pi -
    \left\{(3\,a_r - 2)\,b^2 + (4\,a_r - 5)\,c\right\}\cs
    \right]\Pi\,\Sigma\,\epsilon^3 + \ob(\epsilon^4),       \\[0.45cm]
b,c = 0: &\ds
    -d\,(\Pi - a_r\,\cs)\,\Pi\,\Sigma\,\epsilon^3 + 
    2\,e\,\frac{3\,a_r - 1}{3}\,\cs^3\,\epsilon^4 +
    f\left[
        (a_r - 2)\,\Pi + (3\,a_r - 1)\,\cs
    \right]\Pi\,\Sigma\,\cs\,\epsilon^5 + \ob(\epsilon^6),
\end{array}
\]
with the corresponding values of the trace parameter are
\[
    a_r = a_{1/5,2/5} = \frac{-1\pm\sqrt{5}}{2}.
\]

For the quadratic H\'enon map, where $f(q) = a\,q + b\,q^2$, one
obtains:
\[
\begin{array}{ll}
\ds r_4:    &\ds 
    b^2\Pi^2 \epsilon^2 +
    b^3(\Pi - \cs)\,\Pi\,\Sigma\,\epsilon^3 +                       
    b^4\left(
        2\,\Pi^3 - \Pi^2\,\cs + \frac{\cs^3}{3}
    \right)\epsilon^4 +                                    
    b^5(\Pi - \cs)\,\Pi^2\,\Sigma\,\epsilon^5 + \ob(\epsilon^6),\\[0.45cm]
\ds r_3:    &\ds
    -b\,\Pi\,\Sigma\,\epsilon
    -b^2\left[
        \Pi^2 - \frac{\cs^2}{2}
    \right]\epsilon^2
    -b^3\left[
        \Pi - \frac{\cs}{2}
    \right]\Pi\,\Sigma\,\epsilon^3 + \ob(\epsilon^4),           \\[0.45cm]
\ds r_6:    &\ds
    \frac{b^2}{2}\,\cs^2\epsilon^2 -
    \frac{b^3}{2}\,\Pi\,\Sigma\,\cs\,\epsilon^3 +
    b^4\left[\Pi^3 + \frac{2}{3}\,\cs^3\right]\epsilon^4 +
    b^5\left[
        \Pi^2 - 3\,\Pi\,\cs + \frac{1}{2}\,\cs^2
    \right]\Pi\,\Sigma\,\epsilon^5 + \ob(\epsilon^6),           \\[0.45cm]
\ds r_5:    &\ds
    b^2\,a_r\,\cs^2\epsilon^2 -
    b^3\left[\Pi - (3\,a_r - 2)\,\cs\right]\Pi\,\Sigma\,\epsilon^3 -
    b^4\left[
        3\,\Pi^3 -
        (7\,a_r + 1)\,\Pi^2\,\cs +
        \frac{2}{3}\,(2\,a_r - 1)\,\cs^3
    \right]\epsilon^4,                                          \\[0.45cm]
    &\ds\quad+\,b^5\left[
        2\,(a_r - 3)\,\Pi^2 +
        3\,(2\,a_r + 1)\,\Pi\,\cs -
        (3\,a_r + 2)\,\cs^2
    \right]\,\Pi\,\Sigma\,\epsilon^5 + \ob(\epsilon^6),
    \quad a_r = a_{1/5,2/5} = \frac{-1\pm\sqrt{5}}{2},          \\[0.45cm]
\ds r_8:    &\ds
    b^2(2\,a_r + 1)\,\cs^2\epsilon^2 +
    2\,b^3(a_r - 3)\,\Pi\,\Sigma\,\cs\,\epsilon^3 +
    \ob(\epsilon^4),\qquad\qquad\qquad\qquad\qquad\,\,\,\,
    a_r = a_{1/8,3/8} = \pm\sqrt{2},                            \\[0.45cm]
\ds r_{10}: &\ds
    b^2(3\,a_r + 2)\,\cs^2\epsilon^2 -
    2\,b^3(a_r + 1)\,\Pi\,\Sigma\,\cs\,\epsilon^3 +
    \ob(\epsilon^4),\qquad\qquad\qquad\qquad\qquad\,\,\,\,
    a_r = a_{1/10,3/10} = \frac{1\pm\sqrt{5}}{2}.
\end{array}
\]
For the simple cubic H\'enon map, where $f(q) = a\,q + c\,q^3$,
the expressions become
\[
\begin{array}{llll}
\ds r_4:    &\ds
    -c\,\left[\Pi^2 - \frac{\cs^2}{2}\right]\epsilon^2 -
    c^2\,\Pi^3\,\epsilon^4 + \ob(\epsilon^6),                   &\qquad
\ds r_{3,6}:&\ds
    \frac{c}{2}\,\cs^2\,\epsilon^2 +
    \frac{c^2}{2}\left[
        \Pi^3 - a_r\,(3\,\Pi^2-\cs^2)\,\cs
    \right]\epsilon^4 + \ob(\epsilon^6),                        \\[0.45cm]
\ds r_{5,10}:&\ds
    \frac{3}{2}\,c\,a_r\,\cs^2\,\epsilon^2 + \ob(\epsilon^4),   &\qquad
\ds r_8:&\ds
    \frac{3}{2}\,c\,a_r\,\cs^2\,\epsilon^2 -
    c^2\,(
        3\,\Pi^2 - 5\,\cs^2
    )\,\cs\,\epsilon^4 + \ob(\epsilon^6).
\end{array}
\]

%

\end{document}